\documentclass[11pt]{article}
\usepackage[preprint]{acl}
\usepackage{times}
\usepackage{latexsym}
\usepackage[T1]{fontenc}
\usepackage[utf8]{inputenc}
\usepackage{microtype}
\usepackage{inconsolata}
\usepackage{graphicx}
\usepackage{subcaption}
\usepackage{booktabs}
\usepackage{multirow}
\usepackage{amsmath}
\usepackage{amssymb}
\usepackage{xcolor}
\usepackage{url}

\title{Looks Right, Works Right: \\
A Project-Level Benchmark for Multi-Screen Mobile App Generation}

\author{
Fan Wu\thanks{%
  Computer Science and Technolog, Harbin Institute of Technology. %
  Email: \texttt{codenobuge@163.com}.%
} \And
Cuiyun Gao\thanks{Corresponding author.} \And
Yiming Huang\footnotemark[1]
\AND
Yang Xiao\thanks{Institute of Information Engineering, Chinese Academy of Sciences.} \And
Yujia Chen\footnotemark[1] \And
Qing Liao\footnotemark[1]
}

\begin{document}
\maketitle

\begin{abstract}
Recent multimodal large language models can convert visual designs directly into executable code, but real mobile products require multiple screenshots to become a buildable codebase with shared components and working navigation. This project-level setting exposes three limits of existing design-to-code benchmarks: they focus on single-page generation rather than complete codebases, cannot evaluate cross-page navigation, and do not measure project-wide maintainability. We introduce \textbf{MobileForge}, the first benchmark for project-level multi-screen mobile app generation, comprising $29$ real mobile apps, $309$ human-reviewed screens, structured page-relationship annotations, and $701$ navigation test specifications. MobileForge supports five-axis evaluation of build, navigation, visual fidelity, code maintainability, and efficiency. We also propose \emph{state-isolated navigation testing} to avoid cascading failures in navigation evaluation and an \emph{anchor-referenced list-wise visual evaluation protocol} to improve visual-judge reliability. Across $174$ end-to-end runs on six frontier multimodal LLMs, current models can build mobile-app projects that compile and reach the correct pages, but interactive navigation remains unreliable and visual fidelity and maintainability still lag. The benchmark and supporting materials are available at \url{https://github.com/anoa12159-hue/mobileforge_eval}.
\end{abstract}

\section{Introduction}
\label{sec:intro}

\textit{Can a multimodal LLM read a folder of mobile-app screenshots and generate the source code for that app?} The idea of turning a designer's screenshots directly into a running application has motivated a decade of research, from \textit{pix2code}'s CNN-LSTM screen-to-DSL pipeline \cite{beltramelli2018pix2code} to multimodal LLMs that take a webpage screenshot as input and generate HTML+CSS in one shot \cite{si2024design2code,laurencon2024websight,gui2024webcode2m}. Recent work on single-page design-to-code with rich design-tool metadata further reflects this continued focus on individual pages \cite{gui2026figma2code}.

A real product, however, is never a single screen. It is a folder of screenshots that must collectively become a buildable codebase with shared components, consistent design tokens, working routing, and a navigable user experience. Existing design-to-code benchmarks fall short in three ways exposed by this project-level setting. \textbf{(1) Single-page focus.} They target one webpage at a time and do not evaluate a multi-screen project as a whole, leaving cross-page coherence and shared-component reuse out of scope. \textbf{(2) No evaluation of interactive navigation.} They cannot test whether the generated app's global routing correctly connects tab transitions and parent--child transitions to the intended targets, even though navigation correctness is a basic functional requirement of a usable app. \textbf{(3) No measure of project-wide code maintainability.} They omit indicators such as component reuse, dead-component prevalence, and cross-page design-token consistency, which directly affect downstream engineering cost when a generated project is handed off to a team.

\begin{figure}[t]
\centering
\includegraphics[width=\linewidth]{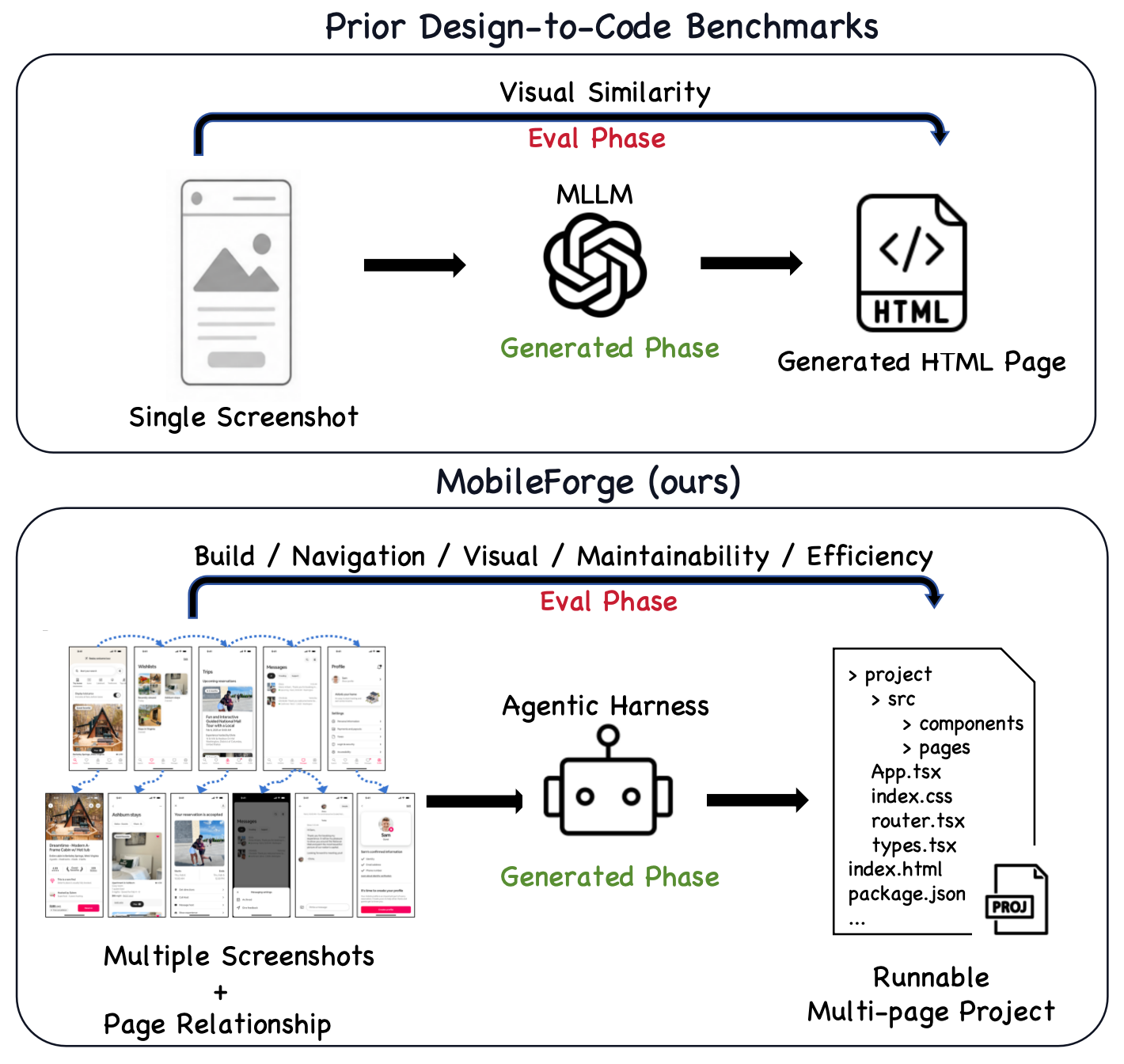}
\caption{Prior design-to-code benchmarks take a \emph{single} screenshot as input and score the resulting HTML page by visual similarity alone (top). MobileForge takes the full set of screenshots of a real mobile app and a page-relationship description as input, asks an agentic harness to produce a runnable multi-page project, and scores it on five orthogonal axes: build, navigation, visual fidelity, code maintainability, and efficiency (bottom).}
\label{fig:intro_case_compare}
\end{figure}

To address these gaps, we introduce \textbf{MobileForge}, the first project-level multi-screen mobile app generation benchmark, together with a five-axis evaluation framework for multi-screen projects (Figure~\ref{fig:intro_case_compare}). Our experiments on six frontier multimodal LLMs across $174$ end-to-end runs show clear gaps between current models and the requirements of project-level interactive code generation, gaps that single-page benchmarks cannot expose.

\paragraph{Key Contributions}
\begin{itemize}
\setlength{\itemsep}{2pt}
\item \textbf{Task and Benchmark.} We formulate project-level multi-screen mobile app generation as a new task and release \textbf{MobileForge}, the first benchmark for this setting, comprising $29$ in-market consumer apps with $309$ human-reviewed screens, structured page-relationship descriptions, and $701$ navigation test specifications.
\item \textbf{Evaluation Framework.} We develop a five-axis evaluation framework covering build, navigation, visual fidelity, code maintainability, and efficiency. Two new evaluation techniques anchor the framework: \emph{state-isolated navigation testing}, which runs each navigation specification from a fixed source-page route rather than a chained walkthrough, and an \emph{anchor-reference list-wise visual evaluation protocol}, calibrated against human raters and motivated by the documented reliability issues of point-wise VLM judging.
\item \textbf{Large-Scale Empirical Study.} We evaluate six frontier multimodal LLMs across $174$ end-to-end runs and find that current models can build mobile-app projects that compile and reach the correct pages, yet interactive navigation is unreliable and visual fidelity and code maintainability leave substantial room for improvement. We release MobileForge, the evaluation harness, and all run artifacts to support reproducibility.
\end{itemize}

\section{Related Work}
\label{sec:related}

\paragraph{Screenshot-to-code and design-to-code.} \textit{Pix2code} \cite{beltramelli2018pix2code} introduced neural code generation from a single GUI screenshot. The modern wave, including \textit{Design2Code} \cite{si2024design2code}, \textit{WebSight} \cite{laurencon2024websight}, \textit{WebCode2M} \cite{gui2024webcode2m}, and recent VLM-centric efforts \cite{ge2025flame,jiang2025screencoder,yang2025ui2coden}, reframes the task as multimodal prompting over web screenshots. However, they still treat each example as a single webpage. \textit{Figma2Code} \cite{gui2026figma2code} studies a complementary setting based on Figma design files, in which structural component definitions are available to the model. A concurrent submission by a partially overlapping author set \cite{anonymous2026webigbench} benchmarks MLLMs on \emph{interactive} single-file HTML/CSS/JavaScript webpages and explicitly scopes out component frameworks, multi-file projects, and multi-page navigation, which are the regimes MobileForge targets. MobileForge differs from all the above in two ways. First, it uses only screenshots and does not assume any structural prior. Second, it targets generation at the project level across multiple screens rather than single page generation.
%Therefore, it is not directly comparable with methods that rely on structural priors or address only single pages in terms of visual perception difficulty.

\paragraph{Mobile UI corpora, agents, and code-generation benchmarks.} A large body of work targets static-UI \emph{perception}, including Rico \cite{deka2017rico} and downstream understanding work \cite{wang2021screen2words,li2020widgetcaptioning,baechler2024screenai,you2024ferretui,hong2024cogagent,lu2024omniparser}, or trains agents to \emph{operate} running apps \cite{zhang2023appagent,wang2024mobileagent,cheng2024seeclick,xie2024osworld}. MobileForge inverts both directions and asks models to \emph{generate} the project. On the code side, benchmarks have moved from function-level (HumanEval \cite{chen2021humaneval}, MBPP \cite{austin2021mbpp}) to repository-level (SWE-bench and variants \cite{jimenez2024swebench,yang2024sweagent,deng2024swebenchpro,tian2026swebenchmobile}) and broader agentic suites \cite{liu2023agentbench,ma2024agentboard}, but none takes a visual design as input. Our single-agent harness draws on the iterative tool-use paradigm \cite{yao2023react,yao2023tot,madaan2023selfrefine,schick2023toolformer,wang2024codeact,xia2025agentless,wang2025openhands,liu2024visualwebbench}; we deliberately fix a single-agent setup so that observed differences are attributable to the model rather than the orchestration \cite{hong2024metagpt,qian2024chatdev,chen2024agentverse,zhang2025maas}.

\paragraph{VLM-as-a-Judge.} Vision-language judges have become standard evaluators when no gold output exists \cite{lee2024prometheusvision,li2026webdevjudge,pairbench2025}, and the literature documents two failure modes of point-wise scoring: scale drift across sessions and ties-in-the-middle compression. Pair-wise and list-wise protocols are more reliable, with list-wise producing a full ranking at $O(N)$ rather than $O(N^2)$ call cost. Our anchor-reference list-wise protocol (\S\ref{sec:visual}) extends list-wise judging with an in-evaluation sanity check, and we substantiate the choice with empirical comparison (Appendix~\ref{app:judge_study}).

\section{The MobileForge Benchmark}
\label{sec:benchmark}

\begin{figure*}[t]
\centering
\includegraphics[width=\linewidth]{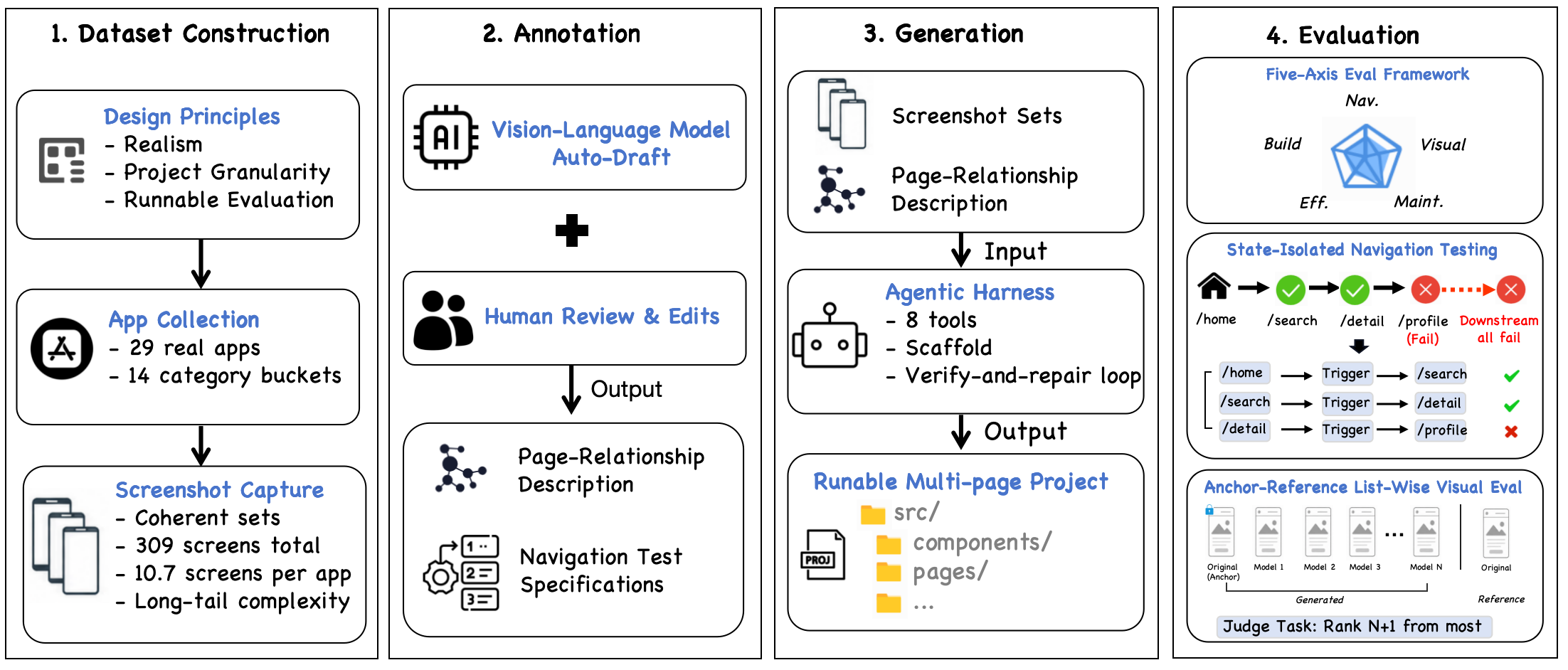}
\caption{MobileForge pipeline. \textbf{(1)~Dataset construction} starts from three design principles: realism, project granularity, and runnable evaluation. It collects $29$ in-market apps across $14$ category buckets with coherent multi-page screenshot sets ($309$ screens, mean $10.7$ per app). \textbf{(2)~Annotation} runs a VLM auto-draft over the screenshots, and then a human reviewer edits the page-relationship description and the navigation test specifications. \textbf{(3)~Generation} hands the reviewed screenshots and page-relationship description to a single-agent harness ($8$ tools, pre-scaffolded directory, verify-repair loop). \textbf{(4)~Evaluation} scores the resulting project on five axes, namely build, navigation, visual fidelity, code maintainability, and efficiency, using state-isolated navigation testing (\S\ref{sec:nav}) and an anchor-reference list-wise visual judge (\S\ref{sec:visual}).}
\label{fig:overview}
\end{figure*}

Figure~\ref{fig:overview} summarizes the four stages of MobileForge: dataset construction, annotation, generation, and five-axis evaluation. The remainder of this section and \S\ref{sec:protocol} describe them in detail.

\subsection{Design Principles}
MobileForge is shaped by three principles: (i) \textbf{realism}, where every instance is an in-market consumer app rather than a synthetic mockup; (ii) \textbf{project granularity}, where each instance is a folder of screens so that cross-page consistency and routing are exercised; and (iii) \textbf{runnable evaluation}, where ground truth is the combination of \emph{building}, \emph{navigating}, and \emph{visually matching}, not pixel-perfect reproduction, which is brittle and underspecified.

\subsection{App Selection and Capture}
We sampled $29$ apps spanning $14$ category buckets (media, finance, social, navigation, e-commerce, productivity, communication, etc.), prioritizing broad cultural recognition. For each app, we captured a coherent screenshot set covering the primary tab structure plus key sub-screens, totaling $309$ screens (mean $10.7$, std $3.8$, range $5$--$21$ per app). The full app list and per-app statistics are in Appendix~\ref{app:per_app} (Table~\ref{tab:per_app_full}).

\subsection{Annotation Schema}
\label{sec:annotation}
Each app's annotation comprises two parts.
The first is a \textbf{page-relationship description} that enters the generation pipeline as model input. It specifies the application's tab structure, namely the main navigation tabs visible at the application root, and the parent--child relationships between screens, encoded as natural-language statements of the form ``\emph{Screen X is a sub-page of Screen Y, triggered by tapping the search icon}.'' This is the structural information the model needs to plan multi-page routing.
The second is a set of \textbf{navigation test specifications} used only at evaluation time. Each specification belongs to one of three categories: \emph{page existence} (is a given screen reachable at the expected route), \emph{tab navigation} (does tapping a tab item route to the correct screen), and \emph{parent--child navigation} (does triggering a known affordance on a parent screen transition to the expected child). Each test records a starting screen, an action with strategy hints (DOM selectors, text matches, tab positions), and the expected target screen. The full benchmark contains $701$ such specifications across $29$ apps (mean $24.2$, range $11$--$41$ per app), distributed as $120$ page-existence, $392$ tab-navigation, and $189$ parent--child cases.

\subsection{Annotation Pipeline and Quality}
\label{sec:annotation_quality}
Annotation follows a two-stage pipeline: a vision-language model drafts the schema entries of \S\ref{sec:annotation}, and human annotators then review every field in a labelling interface (Figure~\ref{fig:annot_tool_page}) and may edit, add, or remove entries before producing the final ground truth. Three properties anchor the dataset's reliability. (i) \emph{Data quality}: every retained annotation has passed human review. (ii) \emph{Annotation-framework accuracy}: against the post-review ground truth, the auto-drafts achieve $96.8\%$ recall and $94.3\%$ precision at the page level, and $74.1\%$ recall and $83.9\%$ precision at the test-case level. (iii) \emph{Low edit rate}: $64.7\%$ of drafted page annotations and $76.5\%$ of test cases are accepted unchanged, confirming that the auto-drafting stage produces candidates that the human reviewer only lightly edits.

\subsection{Benchmark Statistics}
The benchmark spans $309$ screens, $701$ test cases, $10.7$ avg.\ screens per app, and a long-tail distribution of complexity (Khan Academy: $5$ screens / $11$ tests; Klook: $21$ screens / $41$ tests). Table~\ref{tab:benchmark_stats} contrasts MobileForge with prior design-to-code benchmarks along the dimensions that matter for project-level evaluation.

\begin{table*}[t]
\centering
\small
\begin{tabular}{llccccc}
\toprule
\multirow{2}{*}{Benchmark} & \multirow{2}{*}{Unit of evaluation} & \multicolumn{5}{c}{Evaluation Axes} \\
\cmidrule(lr){3-7}
 &  & Build & Nav. & Visual & Maint. & Eff. \\
\midrule
Design2Code \cite{si2024design2code}  & $484$ real webpages              & -- & -- & \checkmark & -- & -- \\
WebSight \cite{laurencon2024websight} & $2$M synth.\ webpages       & -- & -- & \checkmark & -- & -- \\
WebCode2M \cite{gui2024webcode2m}     & $2.56$M real webpages       & -- & -- & \checkmark & -- & -- \\
Figma2Code \cite{gui2026figma2code}   & real Figma files            & -- & -- & \checkmark & \checkmark & -- \\
\textbf{MobileForge} (ours)            & $29$ real apps $/$ $309$ screens & \textbf{\checkmark} & \textbf{\checkmark} & \textbf{\checkmark} & \textbf{\checkmark} & \textbf{\checkmark} \\
\bottomrule
\end{tabular}
\caption{Positioning of MobileForge among design-to-code benchmarks. Prior work evaluates single webpages, primarily along visual similarity; MobileForge introduces project-level evaluation along five axes. Inputs are screenshots throughout, except Figma2Code, which takes Figma metadata. \checkmark indicates that the axis is evaluated.}
\label{tab:benchmark_stats}
\end{table*}

\section{Evaluation Protocol}
\label{sec:protocol}

\subsection{The Five Axes}
\label{sec:axes}

We evaluate each project along five axes. Three use standard measurements, summarized below, and two rest on the new evaluation techniques of \S\ref{sec:nav}--\S\ref{sec:visual}.

\textbf{Build Success.} Each project is shipped to a clean Vite scaffold (React~$19$ + TypeScript + Tailwind). We run \texttt{npx tsc --noEmit} and \texttt{vite build}; a run \emph{builds} only if both pass with zero errors. We additionally report route count, page-file count, and route coverage.

\textbf{Navigation Correctness.} \texttt{NavPassRate}: the fraction of the $701$ navigation specifications (page-existence, tab-navigation, parent--child) passed under state-isolated execution (\S\ref{sec:nav}).

\textbf{Visual Fidelity.} A per-model Borda score in $[0,1]$ aggregated from anchor-reference list-wise ranking by a fixed vision-language judge; a $4$-dimension point-wise rubric is retained as an auditable reference baseline (\S\ref{sec:visual}).

\textbf{Code Maintainability.} Five indicators capture post-generation engineering upkeep: total \texttt{LoC}; mean \texttt{LoC/file} (conciseness); \texttt{ReuseRate}, the average import count per shared component; \texttt{DeadCompRate}, the fraction of declared components that are never imported; and \texttt{ColorConsistency}, the fraction of Tailwind color tokens from a small recurring palette versus one-off shades. We use the term \emph{code maintainability} rather than \emph{code quality} because the former clearly excludes correctness, which is already captured by Build, and keeps the indicators focused on engineering handoff.

\textbf{Efficiency.} For each run, we log input and output tokens, LLM call count, wall-clock LLM time, and a dollar-cost estimate from the model card list price. Efficiency is reported as a first-class axis rather than a footnote, since cost--quality trade-offs surface a distinct finding (\S\ref{sec:frontier}).

\subsection{State-Isolated Navigation Testing}
\label{sec:nav}
Evaluating multi-screen navigation under the standard chained-walkthrough protocol suffers from \emph{error cascading}. Under this protocol, the test agent navigates from the home screen and exercises the entire app as a sequence of clicks. If the bottom-tab routing is broken at step~$3$, every downstream test on the same chain fails too, and the cause of each later failure becomes ambiguous: is it the screen itself, or the upstream tab? In a project with $20$+ test cases per app, a single early break can invalidate half of the chain. This protocol cannot tell us \emph{which} failure mode dominates, and that is precisely the question we want to answer.

We propose \textbf{state-isolated navigation testing}: each test case begins from a \emph{fixed source-page route} provided to the Playwright driver before the action is executed, rather than from a chained sequence starting at the home screen. Concretely, the driver navigates the browser directly to the route of the source page (e.g., \texttt{/search} rather than \texttt{/}~$\rightarrow$~\texttt{/search}), waits for content to settle, executes the action under test, and observes the post-action page. The starting state of every test is therefore independent of every other test.

State isolation produces three benefits. (i) \emph{Statistical independence}: failures are independent events, so per-model failure rates can be compared without inflating denominators on chained failure cascades. (ii) \emph{Fine-grained attribution}: a failure observed on test $N$ is genuinely caused by interaction $N$, which is the prerequisite for the C1--C4 user-perception failure taxonomy we develop in \S\ref{sec:analysis}. (iii) \emph{Reachability decoupled from operability}: tests of the form ``does the target page render at its route?'' (page-existence cases) are separated from tests of the form ``does the trigger fire?'' (parent--child and tab-navigation cases), so reachability and operability can be diagnosed separately.

We render the built project in a headless Playwright browser ($375{\times}812$ viewport, simulating an iPhone~X) and execute the $701$ test specifications. A target is reached if the URL matches the expected page or, for same-route overlays, the page content hash changes meaningfully.

\subsection{Anchor-Reference List-Wise Visual Evaluation}
\label{sec:visual}
Cross-model visual evaluation is structurally hard for VLM-as-Judge. Point-wise scoring, the default in prior design-to-code work, asks the judge to assign a $1$--$5$ score to a single candidate, ignoring how it compares with the alternatives. This produces two well-documented failure modes: \emph{scale drift} across sessions, where the same image receives different scores at different times, and \emph{ties-in-the-middle} compression, where close candidates collapse to the same integer score. Pair-wise judging adds a relative anchor but costs $O(N^2)$ comparisons per scenario, where $N$ is the number of candidate models compared per scenario. Our empirical study (Appendix~\ref{app:judge_study}) further finds that it suffers from $23.3\%$ position bias that requires swap-augmented re-runs. Standard list-wise judging is cheaper ($O(N)$ calls) and produces a full ranking, but it provides no signal on whether the judge actually performed the comparison or simply guessed.

We add a small but consequential modification: we mix the original design screenshot itself, anonymized, into the candidate set as a hidden anchor. The judge ranks $N{+}1$ candidates ($N$ model outputs plus the anchor) from most to least visually faithful. A ranking is valid only when the anchor is placed first, since the anchor is pixel-identical to the reference and any judge that correctly sees the candidates must rank it first. If the anchor is misplaced, the ranking is discarded and the call is re-run.

This anchor mechanism functions on two levels.

\textbf{(i) In-evaluation sanity check.} Every individual judge call is self-validating, and we discard rankings that fail anchor placement rather than including them in the Borda aggregation. The per-call anchor pass rate is itself a quality indicator we report alongside the Borda scores.

\textbf{(ii) Judge-model selection criterion.} The same mechanism becomes a selection criterion for VLM judges. A candidate judge that cannot consistently place the anchor first on a screening set is not eligible to score the benchmark. In our experiments, Gemini~2.5 Pro achieves $\sim\!100\%$ anchor pass on a $33$-scenario screening set; Claude Sonnet~4.5 also achieves $\sim\!100\%$; GPT-5.1 fails at $\sim\!35\%$ because it explicitly reasons that the reference ``is not an independent candidate'' and ranks it last, disqualifying it under this criterion. Appendix~\ref{app:judge_consistency} reports the full cross-judge robustness analysis and a methodological note on the GPT-5.1 failure.

The judge produces a per-scenario ranking, which we aggregate to a per-model Borda score in $[0,1]$ via the $701$-scenario corpus. Borda is the primary visual metric. We additionally retain the per-pair point-wise protocol, a $4$-dimension rubric (Layout $30\%$, Style $25\%$, Content $25\%$, Consistency $20\%$, each $1$--$5$), as an auditable reference baseline. This baseline supplies the absolute-ceiling reading (``no model exceeds $3.0/5$'') that Borda's ordinal scale cannot. Full justification combining theoretical analysis, literature support, and the three-way (point-wise / pair-wise / list-wise) empirical comparison is in Appendices~\ref{app:judge_study} and \ref{app:judge_consistency}.

\section{Experimental Setup}
\label{sec:setup}

\subsection{Agentic Harness}
We deliberately fix the prompting strategy across models to isolate model capability rather than prompt engineering. All models share an identical \textit{agentic} pipeline: a single tool-augmented agent receives the design screenshots and the page-relationship description from \S\ref{sec:annotation}, and a system prompt instructs it to produce a React+TypeScript+Tailwind project under a pre-scaffolded directory. The agent has eight tools: \texttt{write\_file}, \texttt{read\_file}, \texttt{str\_replace}, \texttt{batch\_str\_replace}, \texttt{list\_files}, \texttt{delete\_file}, \texttt{run\_command} (for \texttt{tsc}/build), and \texttt{task\_done}. The agent follows a soft six-phase plan (analyse, foundation, components, pages, assembly, verify) and may iterate up to $50$ times. On \texttt{task\_done}, the harness triggers an automated \texttt{tsc}+\texttt{vite} build; build errors are returned to the agent, which may continue iterating. Smaller models reach the same endpoint by writing files in many small edits and self-repairing more often; the per-phase token attribution accounts for $18$--$35\%$ of total tokens in the verify phase, with the smaller models at the high end.

\subsection{Models}
We evaluate six closed-weight frontier LLMs available with multimodal input as of April~2026: Claude Opus~4.6 and Claude Haiku~4.5 (Anthropic), GPT-5 and GPT-5 Mini (OpenAI), and Gemini~2.5 Pro and Gemini~2.5 Flash (Google). For each app--model cell, we run a single trial, totalling $29\times6=174$ runs, all of which completed and were evaluated end-to-end.

\subsection{Judge Selection}
The vision-language judge for visual fidelity is fixed across all runs as Gemini~2.5 Pro. The selection criterion is that the candidate judge must reliably pass the anchor sanity check on a screening set: when the reference design is mixed into the candidate ranking, a competent judge places it first. Models that fail this check are not eligible. A cross-judge robustness check with Claude Sonnet~4.5 is reported in Appendix~\ref{app:judge_consistency}.

\section{Results}
\label{sec:results}

\subsection{Visual Fidelity Has Not Saturated}
\label{sec:visual_sat}
Table~\ref{tab:overall} reports per-model means across all $29$ apps. List-wise Borda is the primary visual metric, and point-wise visual scores are reported alongside as an auditable absolute-ceiling reference.

\begin{table*}[t]
\centering
\small
\begin{tabular}{lrrrrrrrr}
\toprule
Model & Tok.(K) & \$ & Calls & Build\% & NavPass$\uparrow$ & Borda$\uparrow$ & Vis(pt)$\uparrow$ & LoC \\
\midrule
Claude Opus 4.6   & $488$     & $2.90$ & $14.7$ & $100$ & $\mathbf{92\%}$ & $\mathbf{0.906}$ & $\mathbf{2.99}$ & $1{,}441$ \\
Claude Haiku 4.5  & $1{,}308$ & $1.40$ & $38.5$ & $100$ & $85\%$ & $0.339$ & $2.37$ & $1{,}290$ \\
GPT-5             & $304$     & $0.50$ & $14.7$ & $100$ & $89\%$ & $0.542$ & $2.64$ & $668$ \\
GPT-5 Mini        & $542$     & $0.15$ & $23.9$ & $100$ & $58\%$ & $0.130$ & $1.60$ & $436$ \\
Gemini 2.5 Pro    & $336$     & $0.64$ & $16.0$ & $100$ & $82\%$ & $0.513$ & $2.56$ & $920$ \\
Gemini 2.5 Flash  & $839$     & $0.33$ & $34.5$ & $100$ & $62\%$ & $0.186$ & $2.00$ & $1{,}176$ \\
\bottomrule
\end{tabular}
\caption{Per-model results aggregated over all $29$ apps ($174$ runs). \texttt{Borda}: list-wise Borda score on the $701$-scenario judge run, normalised to $[0,1]$ and used as the primary visual metric. \texttt{Vis(pt)}: point-wise visual mean (1--5 scale; reference baseline). \texttt{Tok.}: input and output token mean. \texttt{Calls}: mean LLM tool-call iteration count. Token differences are dominated by iteration count: Haiku averages $38.5$ iterations versus $14.7$ for Opus and GPT-5, with smaller models writing many small edits and self-repairing more often. Best per column in \textbf{bold}.}
\label{tab:overall}
\end{table*}

Build success has saturated at $100\%$, but visual fidelity has not. List-wise judging shows substantial between-model differences: Claude Opus~4.6 ranks first in $26$ of $29$ apps, the bottom model wins zero apps, and the global Borda scores span the full $[0.13, 0.91]$ range. Point-wise scoring further shows that even the leader does not approach absolute visual faithfulness: the top mean is $2.99/5$ (Opus), with all six models falling in $[1.60, 2.99]$. The gap is concentrated in the cross-page Consistency dimension. Models routinely re-implement navigation chrome inline on each page rather than sharing a single component, which produces subtle color and sizing drift across screens. Cross-page consistency, not single-screen rendering, is the dominant remaining bottleneck, and it is invisible to single-page benchmarks.

\subsection{Code Maintainability Is Orthogonal to Visual Fidelity}
\label{sec:engineering}
Table~\ref{tab:maintainability} reveals a sharp dissociation between visual leadership and code maintainability leadership. Claude Opus~4.6, the visual leader, writes $2.2\times$ the lines of code of GPT-5 ($1{,}441$ versus $668$) at $2.5\times$ the average file size ($82$ versus $33$ LoC/file). GPT-5 also achieves higher component reuse ($4.23$ versus $3.83$ imports per shared component) and $2.5\times$ fewer dead components ($7.8\%$ versus $19.9\%$). Design-token color consistency has largely saturated across all models ($0.74$--$0.85$, a $15\%$ spread), whereas the structural engineering metrics differentiate sharply.

\begin{table*}[t]
\centering
\small
\begin{tabular}{lrrrrr}
\toprule
Model & LoC & LoC/file & Reuse$\uparrow$ & Dead\%$\downarrow$ & Color$\uparrow$ \\
\midrule
Claude Opus 4.6  & $1{,}441$ & $82$ & $3.83$ & $19.9$ & $0.81$ \\
Claude Haiku 4.5 & $1{,}290$ & $68$ & $3.04$ & $12.9$ & $0.85$ \\
GPT-5            & $668$     & $33$ & $\mathbf{4.23}$ & $\mathbf{7.8}$ & $0.80$ \\
GPT-5 Mini       & $436$     & $26$ & $3.49$ & $12.8$ & $0.74$ \\
Gemini 2.5 Pro   & $920$     & $41$ & $2.98$ & $21.4$ & $0.80$ \\
Gemini 2.5 Flash & $1{,}176$ & $52$ & $2.84$ & $17.6$ & $0.78$ \\
\bottomrule
\end{tabular}
\caption{Code maintainability metrics per model (means over $29$ apps). \texttt{LoC/file}: mean lines per source file, used as a conciseness proxy. \texttt{Reuse}: mean import count per shared component. \texttt{Dead\%}: fraction of declared shared components never imported. \texttt{Color}: fraction of Tailwind color tokens from a small recurring palette. Full per-model breakdown including file count, shared-component count, and type utilization is in Appendix~\ref{app:engineering}.}
\label{tab:maintainability}
\end{table*}

The visual leader ranks third on component reuse, fifth on dead-component rate, and last on code conciseness. A team optimizing for designer fidelity at hand-off would choose Opus; a team optimizing for downstream code maintenance would choose GPT-5. Existing single-axis benchmarks conflate these two operating points.

\subsection{Capability-Cost Pareto Frontier}
\label{sec:frontier}
\begin{figure}[t]
\centering
\includegraphics[width=0.95\linewidth]{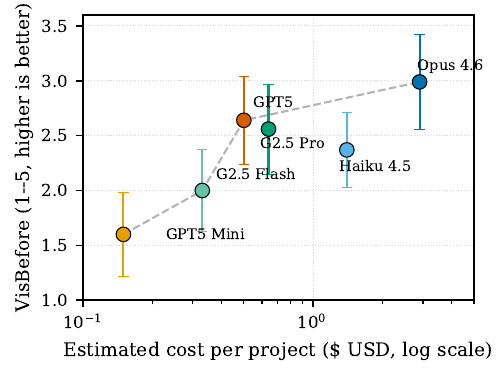}
\caption{Capability-cost frontier on MobileForge. Each point is a model's mean list-wise Borda score versus mean dollar cost across $29$ apps; error bars are $\pm 1$ standard deviation across apps.}
\label{fig:frontier}
\end{figure}

Plotting list-wise Borda against per-app dollar cost (Figure~\ref{fig:frontier}) yields a Pareto frontier where \textbf{GPT-5 dominates two mid-tier models on cost and quality jointly}. GPT-5 (\$0.50/app, Borda $0.542$, NavPass $89\%$) sits strictly inside the dominance region of both Claude Haiku~4.5 (\$1.40, Borda $0.339$, NavPass $85\%$) and Gemini~2.5 Pro (\$0.64, Borda $0.513$, NavPass $82\%$). The $19\times$ cost spread between the cheapest (\$0.15) and most expensive (\$2.90) model does not translate into a proportional quality spread in the relative ranking, and two mid-tier models are dominated. We also observe content fabrication, where models fill otherwise complete UI slots with Lorem ipsum text or stock-photo imagery instead of the screenshot's actual content, predominantly in the lowest-cost models (GPT-5 Mini and Gemini~2.5 Flash); we treat this as part of the cost-quality picture rather than as a separate navigation failure mode.

\subsection{Per-App Variance}
\label{sec:per_app}
Difficulty varies sharply across apps. The easiest apps, Notion ($3.62$), Khan Academy ($3.50$), and WhatsApp ($3.33$), are visually clean with compact tab structures. The hardest apps, Apple News ($2.01$), Airbnb ($2.02$), and Coinbase ($2.07$), combine dense data presentation with custom typography and information-rich imagery. Klook ($21$ screens, $0.48$ NavPass) and Medium ($10$ screens, $0.45$ NavPass) illustrate the cross-page consistency penalty: many screens magnify per-screen weaknesses and stress component-sharing decisions. The full per-app table is in Appendix~\ref{app:per_app}; we read this distribution as evidence that the $29$-app pool covers a wide spread of project-level difficulty.

\section{Failure Analysis}
\label{sec:analysis}

We focus our failure analysis on interactive navigation, namely the $581$ test specifications for tab navigation and parent--child navigation, which directly test the model's overall routing structure. Across all $174$ runs ($6$ models $\times$ $29$ apps), these specifications generate $3{,}486$ test-case executions. We hand-label a stratified sample of $64$ failures under a four-class user-perception schema (C1--C4) and project the sample proportions to the population with Wilson $95\%$ confidence intervals. State-isolated testing (\S\ref{sec:nav}) makes this fine-grained attribution possible: a failure observed on a given test cannot be blamed on an upstream chained interaction.

Three modes account for roughly $90\%$ of interactive navigation failures (Table~\ref{tab:failures}).

\begin{table}[t]
\centering
\small
\setlength{\tabcolsep}{4pt}
\begin{tabular}{llrr}
\toprule
Code & Description & $\hat{p}$ & $95\%$ CI \\
\midrule
C1 & Blank starting page & $7.8\%$ & $[3.0, 18.3]$ \\
C2 & Route mapping error & $10.9\%$ & $[5.4, 20.9]$ \\
C3 & Target unreachable & $32.8\%$ & $[22.6, 45.0]$ \\
C4 & Click has no effect & $48.4\%$ & $[36.6, 60.4]$ \\
\bottomrule
\end{tabular}
\caption{Distribution of interactive navigation failure modes from a stratified sample of $n{=}64$.}
\label{tab:failures}
\end{table}
%. Wilson $95\%$ CI.

\paragraph{C1: blank starting page ($7.8\%$).} The source page renders empty or as a stub placeholder without substantive content or interaction targets, so no navigation action can be dispatched (Figures~\ref{fig:case_C1_1}--\ref{fig:case_C1_2}). This is distinct from the filled-but-fake content fabrication discussed in \S\ref{sec:frontier}: in C1 the relevant affordances are absent, whereas in fabrication the UI structure is present but its text or imagery is placeholder content. C1 is the distinctive failure mode for Gemini~2.5 Flash, which alone accounts for an outsized share of C1 occurrences: $42\%$ of its labelled failures are C1, versus $0\%$ for the four next-best models. Flash's high iteration count ($34.5$ on average) does not translate into substantive page content for failed cases.

\paragraph{C2: route mapping error ($10.9\%$).} The agent declared the route, but the URL resolves to a different page than the one referenced by the test (Figure~\ref{fig:case_C2}). The model has correctly produced the page; the global routing table simply maps the navigation action to the wrong page.

\paragraph{C3: target unreachable or occluded ($32.8\%$).} The action target is not present in the rendered viewport, either because the layout omits it entirely or because another element covers it (Figures~\ref{fig:case_C3_1}, \ref{fig:case_C3_2}, \ref{fig:case_C3_3}). This is structurally distinct from C4: in C3 the page is malformed, whereas in C4 it is well-formed but inert. C3 dominates the smaller models, where layout sparsity is the recurring failure mode.

\paragraph{C4: target clickable but unresponsive ($48.4\%$).} The page renders, the affordance is in the right place, but the click registers no effect (Figures~\ref{fig:case_C4_1}, \ref{fig:case_C4_2}, \ref{fig:case_C4_3}). This is consistent with a missing or mis-wired \texttt{onClick} handler. Build verification catches type errors but not silent event-handler bugs; this is the dominant failure mode for the stronger models.

\paragraph{Per-model headline observations.} The failure-mode distribution differs sharply across models (per-model $n{=}8$--$12$; we read the per-model proportions as descriptive signatures rather than precise point estimates). Claude Opus~4.6 and GPT-5 show identical signatures: $75\%$ C4 in our labelled sample (Table~\ref{tab:per_model_failures}), indicating that their failures concentrate on broken interaction wiring rather than missing UI. Gemini~2.5 Pro is similar but less extreme ($55\%$ C4, $45\%$ C3). GPT-5 Mini reverses the pattern: $67\%$ C3, so when GPT-5 Mini fails, it most often produces a layout that simply lacks the target affordance. Gemini~2.5 Flash is the C1 outlier ($42\%$). Claude Haiku~4.5's failures are more uniformly spread across C2--C4, consistent with its higher route-mapping error contribution. Practically, the remaining failures are not subtle: handlers that do not fire and targets that are not drawn, both diagnosable from the generated component tree before runtime. Full per-model and per-scenario-type tables are in Tables~\ref{tab:per_model_failures} and~\ref{tab:per_type_failures}.

\section{Conclusion}
\label{sec:conclusion}

We introduce MobileForge, the first project-level benchmark for multi-screen mobile app generation, and evaluate six frontier LLMs end-to-end on $174$ runs along a five-axis protocol. The answer to our opening question is \emph{partly yes}: every model produces a buildable project and the strongest reaches $92\%$ navigation correctness, yet visual fidelity has not saturated, maintainability is uncorrelated with visual fidelity, and roughly $90\%$ of navigation failures concentrate in three predictable modes.

\section*{Limitations}
\label{sec:limitations}

\paragraph{Single trial per cell.} Each (app, model) pair runs once. We chose $29\times 6$ apps over fewer apps with more trials to maximise diversity; a multi-trial replication is left to future work.

\paragraph{Vision-language judge.} Visual scores are produced by Gemini~2.5 Pro, a competitor model class. We fix the judge across all runs, so all models share the same judge bias, calibrate the judge against human ratings on $55$ scenarios (strict Kendall $\tau{=}0.716$, top-1 accuracy $83.6\%$), and cross-check with Claude Sonnet~4.5 on $139$ scenarios (inter-judge $\tau{=}0.867$). No self-preference bias toward either judge family is observed.

\paragraph{Web-rendering proxy.} The output target is web (React+Tailwind), not native iOS/Android. We chose web for evaluability, including browser automation and deterministic builds, at the cost of not exercising native APIs. Cross-platform extension is a natural follow-up.

\section*{Ethical Considerations}

\paragraph{Data sources and licensing.} The $29$ apps in MobileForge are widely distributed consumer apps obtained through their official public distribution channels, and screenshots were captured manually from default-state or demo-mode screens without bypassing paywalls, login walls, or rate-limit protections. To respect the intellectual property of the source apps, our public release contains the captured screenshot set together with the structured page-relationship annotations and navigation test specifications, framed as a research benchmark rather than a verbatim re-distribution of any app's UI assets. The benchmark and the accompanying harness and evaluator are released under a research-only license that prohibits commercial reuse of the screenshot artifacts and any derivative product that re-skins them.

\paragraph{Privacy and PII.} MobileForge focuses on UI-level structure and navigation on the public surfaces of consumer apps. Annotators captured screens without logging into real user accounts, and no personal profiles, payment information, private messages, or location histories were accessed or collected. We manually inspected every screenshot and redacted incidental content that could disclose third-party PII before release.

\paragraph{Human annotators.} The page-relationship descriptions and navigation test specifications were authored by members of our research team rather than crowdworkers. Annotators were fully informed about the purpose and intended release of the dataset, participated voluntarily, and were compensated above the local minimum wage standard of their region.

\paragraph{Use of pre-trained LLMs and compute.} Our experiments rely on publicly available commercial LLM inference APIs and a publicly available VLM judge. API usage conforms to the respective providers' Terms of Service. Because agentic code generation with frontier LLMs is energy-intensive, our pipeline runs each (app, model) cell exactly once and caches all intermediate artifacts so that downstream metric recomputation does not re-trigger generation or browser automation. Token consumption and wall-clock time per app are reported transparently as part of the efficiency axis (\S\ref{sec:axes}).

\paragraph{Potential misuse and mitigation.} A capable design-to-code model could in principle be misused to clone legitimate apps for phishing, impersonation, or counterfeit distribution. We mitigate this risk in several ways. MobileForge is a benchmark, not a generator or a fine-tuning corpus, and it does not lower the barrier to producing malicious clones beyond what off-the-shelf multimodal LLMs already provide. The five-axis evaluation rewards build, navigation correctness, visual fidelity, maintainability, and efficiency, none of which is informative for malicious intent. The research-only license further forbids commercial use, ruling out the most plausible misuse vectors such as deploying re-skins of the source apps for profit.

\paragraph{AI writing and coding assistance.} Per the ACL policy on AI writing assistance, we disclose that general-purpose AI assistants were used solely for language polishing and for low-level coding assistance during the implementation of the agentic harness and the evaluator. All scientific claims, experimental design, data analysis, and conclusions are the work of the human authors, who take full responsibility for the correctness of the content.

\section*{Reproducibility Statement}
The benchmark, the agentic harness, and the five-axis evaluator are released as a single repository, together with the $174$ run artifacts. Each run includes the generated source tree, the build log, the navigation-test screenshots, the VLM judge rationales, and a structured run-summary record. A one-command replay reproduces the comparison table verbatim.

\bibliography{references}

\appendix

\section{Discussion Note}
\label{app:method_note}

\paragraph{Build success is a property of the harness, not the model.} The verify-repair loop with \texttt{tsc} and \texttt{vite} lets any frontier model iterate until the project compiles, so the discriminating signal moves to visual fidelity (Borda spread $[0.13, 0.91]$) and navigation correctness ($58\%$--$92\%$). Benchmarks that want compilation as a discriminator must remove the verify-repair loop or impose stricter gates, such as no \texttt{any}, strict null checks, and a lint pass.

\section{Per-App Results}
\label{app:per_app}

Table~\ref{tab:per_app_full} reports per-app means across all six models. Table~\ref{tab:per_app_borda_kendall} summarizes the per-app list-wise Borda leader and the agreement between per-app and global model rankings.

\begin{table}[h]
\centering
\small
\begin{tabular}{lrrr}
\toprule
App & Pages & NavPass$\uparrow$ & Vis(pt)$\uparrow$ \\
\midrule
notion & $4.8$ & $0.90$ & $3.62$ \\
khan-academy & $5.0$ & $1.00$ & $3.50$ \\
whats-app & $15.5$ & $0.84$ & $3.33$ \\
shopify & $9.2$ & $0.98$ & $3.29$ \\
google & $7.3$ & $0.77$ & $3.16$ \\
you-tube-music & $7.8$ & $0.74$ & $3.16$ \\
revolut & $14.2$ & $0.88$ & $3.12$ \\
uber & $8.3$ & $0.77$ & $3.07$ \\
netflix & $5.5$ & $0.79$ & $3.04$ \\
wise & $6.5$ & $0.94$ & $3.03$ \\
pinterest & $12.0$ & $0.70$ & $2.98$ \\
reddit & $6.5$ & $0.83$ & $2.98$ \\
messenger & $5.5$ & $0.60$ & $2.96$ \\
medium & $10.3$ & $0.45$ & $2.94$ \\
coursera & $8.8$ & $0.89$ & $2.93$ \\
grab & $14.2$ & $0.72$ & $2.91$ \\
klook & $16.7$ & $0.48$ & $2.82$ \\
fiverr & $15.0$ & $0.77$ & $2.81$ \\
spotify & $5.2$ & $0.86$ & $2.80$ \\
naver & $7.8$ & $0.48$ & $2.74$ \\
tik-tok & $7.0$ & $0.87$ & $2.74$ \\
uber-eats & $11.2$ & $0.92$ & $2.69$ \\
discord & $11.2$ & $0.80$ & $2.56$ \\
etsy & $10.7$ & $0.82$ & $2.56$ \\
x & $8.0$ & $0.76$ & $2.47$ \\
nextdoor & $8.7$ & $0.69$ & $2.43$ \\
coinbase & $8.8$ & $0.76$ & $2.07$ \\
airbnb & $10.0$ & $0.80$ & $2.02$ \\
apple-news & $9.0$ & $0.78$ & $2.01$ \\
\bottomrule
\end{tabular}
\caption{Per-app means across all six models ($n{=}6$ per row). Sorted by \texttt{Vis(pt)} descending.}
\label{tab:per_app_full}
\end{table}

\begin{table}[h]
\centering
\small
\setlength{\tabcolsep}{4pt}
\begin{tabular}{lcclr}
\toprule
App & Top-1 & Borda & Runner-up (Borda) & $\tau_{\text{glob}}$ \\
\midrule
\multicolumn{5}{l}{\emph{Three apps where Opus does not lead}} \\
etsy           & GPT-5 & $0.886$ & Opus ($0.787$)    & $0.47$ \\
pinterest      & GPT-5 & $0.716$ & Opus ($0.691$)    & $0.47$ \\
youtube-music  & GPT-5 & $0.632$ & Opus ($0.566$)    & --     \\
\midrule
\multicolumn{5}{l}{\emph{Selected Opus wins by Borda}} \\
khan-academy & Opus & $1.000$ & Haiku ($0.583$)   & $0.20$ \\
reddit       & Opus & $1.000$ & GPT-5 ($0.667$)   & $0.73$ \\
grab         & Opus & $0.992$ & Gem-Pro ($0.610$) & $0.73$ \\
airbnb       & Opus & $0.982$ & GPT-5 ($0.535$)   & $0.87$ \\
klook        & Opus & $0.981$ & GPT-5 ($0.222$)   & $0.87$ \\
coinbase     & Opus & $0.974$ & GPT-5 ($0.645$)   & $0.73$ \\
naver        & Opus & $0.877$ & GPT-5 ($0.636$)   & $0.33$ \\
nextdoor     & Opus & $0.703$ & Gem-Pro ($0.667$) & $0.60$ \\
\bottomrule
\end{tabular}
\caption{Per-app list-wise Borda leader, runner-up, and the Kendall $\tau$ between the app's $6$-model ranking and the global ranking. The $26$ remaining Opus-win apps follow the same global ranking pattern and are omitted for brevity (per-app $\tau_{\text{glob}}$ ranges from $0.20$ to $1.0$, mean $0.715$). GPT-5 takes Etsy and Pinterest, which are image-heavy catalogue layouts where Opus's text-fidelity advantage is less decisive, as well as YouTube Music. The two lowest per-app $\tau_{\text{glob}}$ values, khan-academy ($0.20$) and naver ($0.33$), are driven by unusual runner-ups (Haiku and GPT-5 respectively) rather than disagreement on the leader.}
\label{tab:per_app_borda_kendall}
\end{table}

\section{Judge-Method Study}
\label{app:judge_study}

\paragraph{Three-way comparison.} On a $3$-scenario $\times$ $5$-model smoke-test, list-wise achieves shuffle-stability Kendall $\tau{=}0.822$ and a $100\%$ anchor pass rate at $9$ API calls per scenario set. Pair-wise exhibits $23.3\%$ position bias, requiring swap-augmented runs and doubling the call count to $60$ API calls. Point-wise produces ``ties-in-the-middle'' on close candidates in every scenario tested. Inter-method Kendall $\tau$ values (pair $\leftrightarrow$ list $0.667$, point $\leftrightarrow$ list $0.667$, point $\leftrightarrow$ pair $0.733$) confirm that all three methods agree on top and bottom but disagree in the mid-range, which is precisely the regime where a benchmark needs discrimination.

\paragraph{Full-scale list-wise.} List-wise with $k$-shuffle$=3$ and a reference anchor was run across all $29$ apps $\times$ $701$ scenarios with Gemini~2.5 Pro as judge. Per-app Borda scores feed Tables~\ref{tab:overall} and \ref{tab:per_app_borda_kendall}. Per-scenario-type rankings (page-existence / parent--child / tab-navigation) align with the global ranking at Kendall $\tau \geq 0.867$.

\paragraph{Human alignment.} We labelled $55$ scenarios with a blinded ranking interface (anonymous candidate labels A, B, C, $\dots$). List-wise versus human: strict Kendall $\tau{=}0.716$, top-$1$ accuracy $83.6\%$, top-$2$ overlap $80.4\%$, and unrankable-set IoU, meaning agreement on which candidates are ``unrankable'', of $0.968$. Page-existence scenarios show $\tau{=}0.699$ and zero unrankable candidates; parent--child scenarios show $\tau{=}0.731$, and $21/32$ scenarios contain at least one unrankable candidate. The unrankable signal is concentrated entirely in parent--child cases, with Gemini~2.5 Flash accounting for $32\%$ of the $62$ unrankable candidates.

\section{Cross-Judge Consistency}
\label{app:judge_consistency}

We assess two robustness properties of the anchor-reference list-wise protocol: (i) whether the per-model Borda ranking is stable when the judge is swapped out, and (ii) whether the anchor sanity check correctly separates eligible from ineligible judge candidates.

\paragraph{Inter-judge agreement.} Re-running the protocol with Claude Sonnet~4.5 as a second judge on $139$ scenarios drawn from $5$ apps yields inter-judge Kendall $\tau{=}0.867$ on the global per-model Borda ranking. The top three and bottom three model positions are identical across the two judges; the only swap is in the middle, where Gemini~2.5 Pro and Claude Haiku~4.5 trade positions. The two judges therefore agree on the substantive findings: Opus leads, the two GPT-5 variants and Gemini~2.5 Pro form a middle cluster, and GPT-5 Mini sits last. They disagree only on an ordering within the middle band that does not affect the main conclusion.

\paragraph{Self-preference bias.} A standard concern with LLM-as-Judge is family-level self-preference: a Claude judge might systematically reward Claude candidates, and similarly for Gemini. We observe the opposite. The Claude Sonnet~4.5 judge gives Claude-family candidates (Opus, Haiku) slightly \emph{lower} normalised Borda than the Gemini~2.5 Pro judge does, while the Gemini judge gives Gemini candidates (Pro, Flash) almost identical Borda to what the Claude judge gives them. The direction of any small effect is opposite to family preference; no judge inflates its own family.

\paragraph{Anchor sanity check as judge eligibility.} The anchor-pass rate functions as a selection criterion for VLM judges. On a $33$-scenario screening set, Gemini~2.5 Pro achieves $\sim\!100\%$ anchor placement, and Claude Sonnet~4.5 also reaches $\sim\!100\%$; both qualify and produce mutually consistent global rankings (above). GPT-5.1, by contrast, places the anchor first on only $\sim\!35\%$ of calls. Inspection of its rationales shows that GPT-5.1 explicitly meta-reasons about the prompt: it identifies the anchor as the reference image and concludes that the reference ``is not an independent candidate'' and should not be ranked alongside the model outputs, often placing it last instead. GPT-5.1 is consequently ineligible under our criterion because its outputs would not be discardable under the anchor-validity rule, and including them would introduce a non-random sampling bias against the protocol. We report this as a methodological observation rather than a model-quality claim: list-wise plus anchor judging depends on judges that follow the prompt's literal ranking instruction, and protocols that rely on the judge to ``do the obvious thing'' may behave differently as judges become more aggressive about meta-interpretation.

\paragraph{Per-scenario-type stability.} Decomposing the global ranking by the three scenario types (page-existence, parent--child navigation, tab navigation) reveals high stability: the rankings for parent--child and tab-navigation align with the global ranking at Kendall $\tau{=}1.0$, and page-existence aligns at $\tau{=}0.867$ (one swap: GPT-5 and Gemini~2.5 Pro). The full per-type Borda matrix is released alongside the benchmark.

\section{Agentic Harness Prompt}
\label{app:prompt}
The system prompt instructs the model to produce a complete React+TypeScript+Tailwind project, lists the eight available tools, and specifies the six-phase plan (analyse, foundation, components, pages, assembly, verify). We deliberately do not include screenshot-specific instructions or chain-of-thought primers \cite{wei2022cot,yao2023react} so that the result is attributable to the model rather than prompt engineering. The complete prompt is in the released repository.

\section{Trajectory and Context Composition}
\label{app:trajectory}

Tables~\ref{tab:tool_mix} and~\ref{tab:ctx_composition} report what the agent actually does across the $174$ runs: the share of each tool in the tool-call stream, and where the input-token budget accumulates in the context window. Both are means over $29$ apps per model.

\paragraph{Tool-call mix.} The dominant differentiator is \emph{how many files} a model writes per \texttt{write\_file} call. Opus, GPT-5, and Gemini~2.5 Pro batch ($3.3$--$8.5$ files per call; batch ratio $0.51$--$0.70$); Claude Haiku~4.5 writes exactly one file per call (batch ratio $0.00$), which mechanically drives its $38.5$-iteration mean. Only the Gemini family uses \texttt{batch\_str\_replace} non-trivially ($5$--$10\%$ of calls). Claude models call \texttt{read\_file} $3$--$10\times$ more often than GPT-5 or Gemini, which we read as a self-verification habit: re-reading written code before subsequent edits. \texttt{run\_command} (build/typecheck) usage is highest for Haiku ($13.2\%$), consistent with the self-repair pattern noted in \S\ref{sec:setup}. The single-edit \texttt{str\_replace} and \texttt{delete\_file} tools are unused by every model in our six-model set and are omitted from the table; all models gravitate to whole-file \texttt{write\_file} edits or, in the Gemini case, batched replacements.

\paragraph{Context composition.} For every model, the two largest contributors to cumulative input are screenshot tokens and the model's own \texttt{write\_file} outputs echoed back as tool results; together they account for $47\%$ (Gemini~2.5 Flash) to $84\%$ (GPT-5) of cumulative input. Read-file results, build logs, and task updates together fall under $8\%$ in every cell, so the apparent ``observation cost'' of the agentic loop is dominated by the model re-reading what it has already written rather than by external feedback. The \emph{Overhead} column captures message-wrapping tokens, including tool-result envelopes, role markers, and function-calling protocol scaffolding, and varies with both iteration count and per-provider protocol verbosity. Together, these factors push Gemini~2.5 Flash's overhead share above $40\%$.

\begin{table*}[ht]
\centering
\small
\begin{tabular}{lrrrrrr}
\toprule
Model & \texttt{write\_file} & \texttt{batch\_str\_replace} & \texttt{read\_file} & \texttt{run\_command} & \texttt{update\_tasks} & Other \\
\midrule
Claude Opus 4.6  & $47.9$ & $0.7$  & $16.3$ & $8.5$  & $20.4$ & $6.3$ \\
Claude Haiku 4.5 & $51.2$ & $1.9$  & $11.1$ & $13.2$ & $17.0$ & $5.2$ \\
GPT-5            & $67.5$ & $0.5$  & $3.6$  & $4.1$  & $17.7$ & $6.5$ \\
GPT-5 Mini       & $65.2$ & $0.4$  & $2.1$  & $7.2$  & $18.0$ & $6.9$ \\
Gemini 2.5 Pro   & $66.2$ & $5.5$  & $1.7$  & $5.3$  & $18.1$ & $3.2$ \\
Gemini 2.5 Flash & $66.3$ & $10.3$ & $2.6$  & $4.4$  & $12.8$ & $3.6$ \\
\bottomrule
\end{tabular}
\caption{Tool-call mix per model (\% of all tool invocations, means over $29$ apps). \texttt{update\_tasks} is the phase-tracker helper that enforces the six-phase plan (separate from the eight content tools listed in \S\ref{sec:setup}). Other = \texttt{list\_files} + \texttt{task\_done}. The single-edit \texttt{str\_replace} and \texttt{delete\_file} are unused by all six models and are omitted. Rows may not sum to $100\%$ due to rounding.}
\label{tab:tool_mix}
\end{table*}

\begin{table*}[ht]
\centering
\small
\begin{tabular}{lrrrrr}
\toprule
Model & Images & Prompt & \texttt{write\_file} echo & Other tool results & Overhead \\
\midrule
Claude Opus 4.6  & $38.6$ & $2.9$ & $37.7$ & $4.3$ & $16.5$ \\
Claude Haiku 4.5 & $39.0$ & $2.9$ & $32.9$ & $7.6$ & $17.6$ \\
GPT-5            & $52.3$ & $5.2$ & $31.5$ & $2.7$ & $8.4$  \\
GPT-5 Mini       & $53.7$ & $5.1$ & $21.1$ & $2.3$ & $17.8$ \\
Gemini 2.5 Pro   & $11.1$ & $4.7$ & $47.9$ & $4.1$ & $32.2$ \\
Gemini 2.5 Flash & $9.0$  & $3.6$ & $38.2$ & $4.3$ & $44.9$ \\
\bottomrule
\end{tabular}
\caption{Context composition per model (\% of cumulative input tokens at end of run, means over $29$ apps). \emph{Images}: API-reported image tokens for the screenshot payload. The same screenshots are sent to all models, but each provider's tokenizer accounts for image content differently. Gemini reports substantially fewer image tokens per screenshot than Claude or GPT-5, so this row reflects tokenizer accounting, not screenshot count. \emph{Prompt}: system prompt plus initial user text. \texttt{write\_file} \emph{echo}: \texttt{write\_file} tool results returned to the agent as context. \emph{Other tool results}: \texttt{read\_file}, \texttt{run\_command}, \texttt{list\_files}, \texttt{update\_tasks}, \texttt{str\_replace}/\texttt{batch\_str\_replace}, and assistant-text tokens combined. \emph{Overhead}: message-wrapping tokens not attributable to any tool. Rows may not sum to $100\%$ due to rounding.}
\label{tab:ctx_composition}
\end{table*}

\section{Annotation Pipeline Detail}
\label{app:annotation}

Table~\ref{tab:annotation_quality} reports the aggregate annotation-quality statistics referenced in \S\ref{sec:annotation_quality}. Per-app review states (all apps passed review) and field-level edit distributions are released alongside the benchmark.

\begin{table}[h]
\centering
\small
\begin{tabular}{lrr}
\toprule
Metric & Pages & Test Cases \\
\midrule
Recall (auto $\rightarrow$ reviewed) & $96.8\%$ & $74.1\%$ \\
Precision (auto $\rightarrow$ reviewed) & $94.3\%$ & $83.9\%$ \\
Unchanged rate (auto correct as-is) & $64.7\%$ & $76.5\%$ \\
Total reviewed & $309$ & $701$ \\
\bottomrule
\end{tabular}
\caption{Annotation pipeline quality. Recall: fraction of final reviewed items already produced by the auto-drafting model. Precision: fraction of auto-drafted items retained in the reviewed set. Unchanged: fraction of auto items retained with no field-level edits.}
\label{tab:annotation_quality}
\end{table}

\begin{figure*}[t]
\centering
\includegraphics[width=0.92\textwidth]{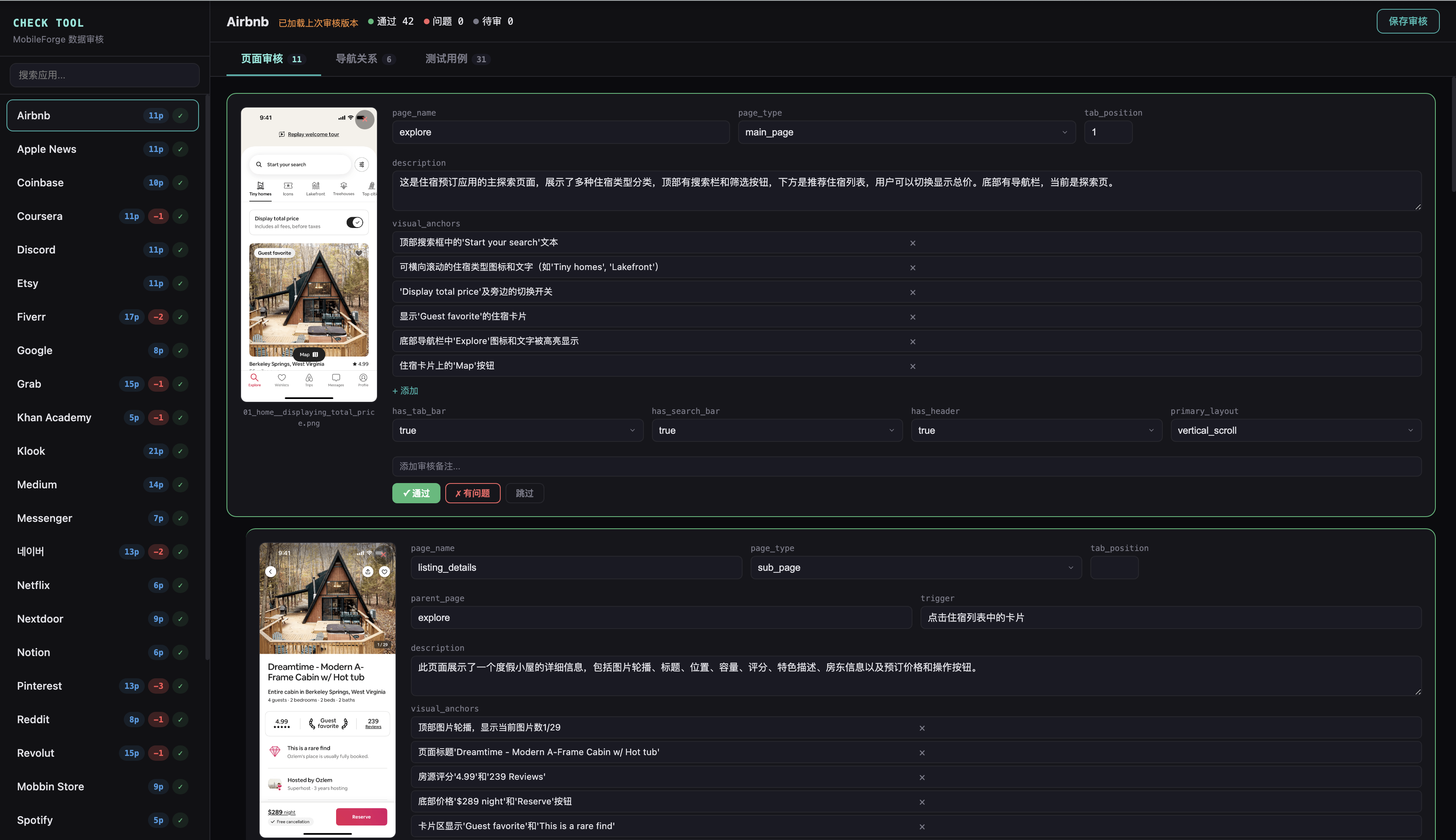}
\caption{Annotation review interface, per-page pane (\S\ref{sec:annotation_quality}). The reviewer inspects each page's type, description, key elements, layout, and primary action, and may pass, edit, or reject the auto-drafted entry.}
\label{fig:annot_tool_page}
\end{figure*}

\begin{figure*}[t]
\centering
\includegraphics[width=0.92\textwidth]{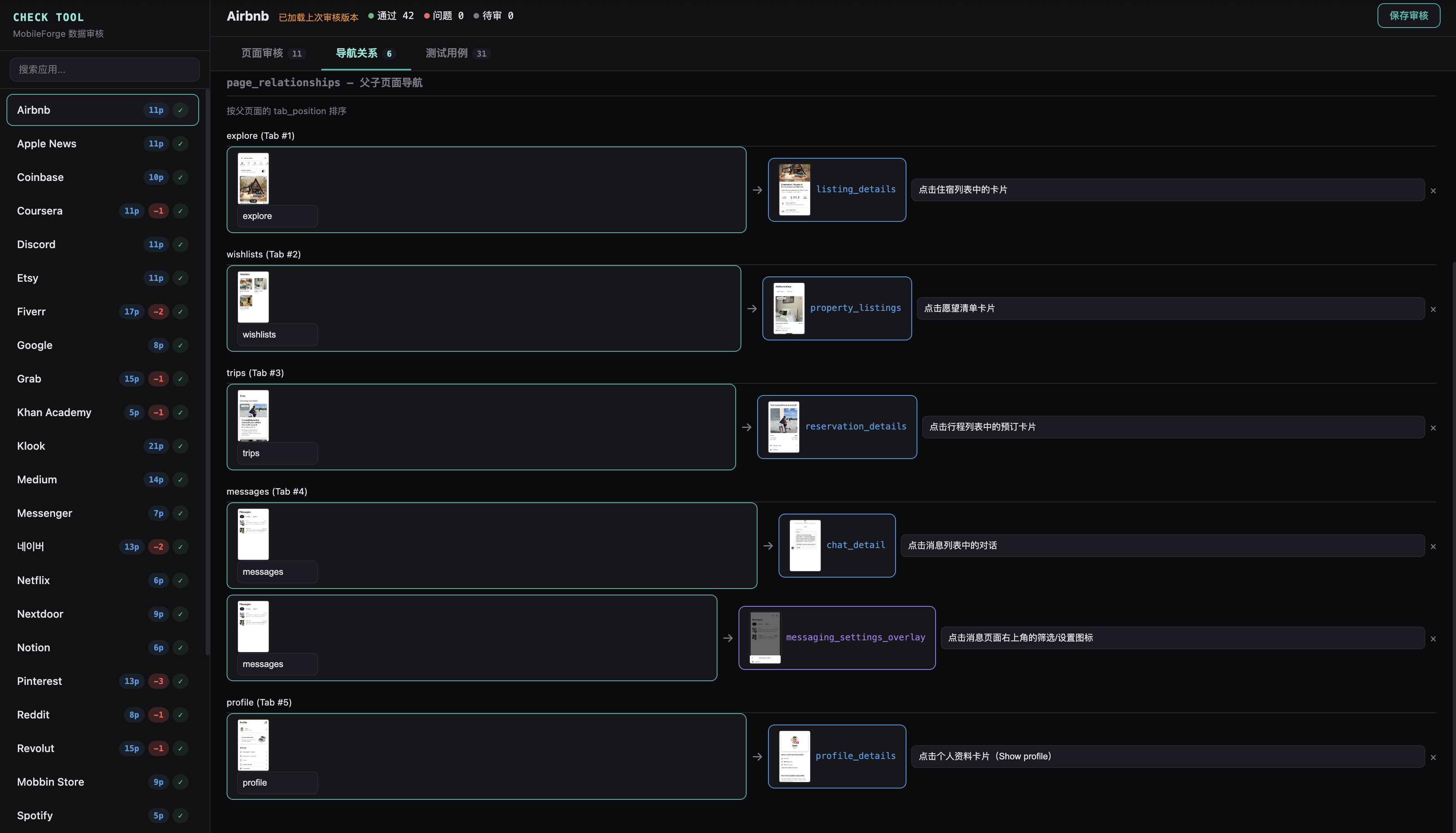}
\caption{Annotation review interface, page-relationship pane. Parent--child and tab pairs are surfaced with source/target screenshots and the trigger description; the reviewer pass/edits/rejects each pair.}
\label{fig:annot_tool_nav}
\end{figure*}

\begin{figure*}[t]
\centering
\includegraphics[width=0.92\textwidth]{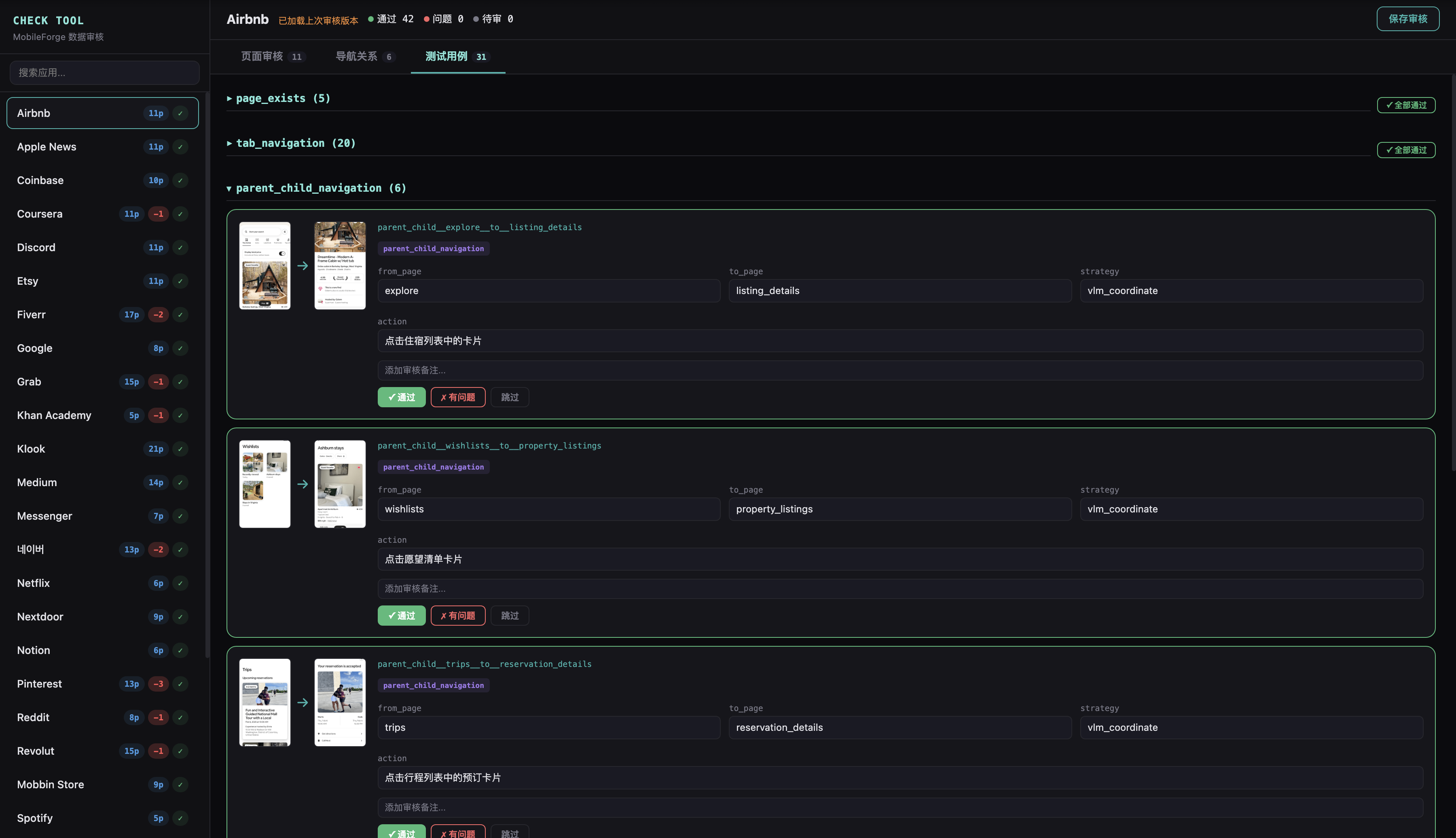}
\caption{Annotation review interface, navigation test-case pane. Page-existence, tab-navigation, and parent--child specifications carry pass / has-issue / reject states aggregated into Table~\ref{tab:annotation_quality}.}
\label{fig:annot_tool_tests}
\end{figure*}

\paragraph{Annotation review interface.} Figures~\ref{fig:annot_tool_page}--\ref{fig:annot_tool_tests} show the three review panes of the in-house annotation tool used to produce the reviewed ground truth. The per-page pane (Figure~\ref{fig:annot_tool_page}) exposes page-level fields, the page-relationship pane (Figure~\ref{fig:annot_tool_nav}) surfaces parent--child and tab pairs, and the test-case pane (Figure~\ref{fig:annot_tool_tests}) carries the navigation specifications. Each pane lets the human reviewer pass, edit, or reject every draft item produced by the auto-drafting stage. The pass/edit/reject states aggregated across all $309$ pages and $701$ test cases produce the statistics in Table~\ref{tab:annotation_quality}.

\section{Failure Taxonomy Detail}
\label{app:failure_taxonomy}

Table~\ref{tab:per_model_failures} shows the per-model C-code distribution within the $64$-sample interactive-navigation failure set (\S\ref{sec:analysis}). Table~\ref{tab:per_type_failures} projects the same labels onto the three scenario types, exploiting the reachability--operability decoupling that state-isolated testing makes possible (\S\ref{sec:nav}).

\begin{table}[h]
\centering
\small
\begin{tabular}{lrrrr}
\toprule
Model & C1 & C2 & C3 & C4 \\
\midrule
Claude Opus 4.6  & $0$  & $17$ & $8$  & $75$ \\
Claude Haiku 4.5 & $0$  & $33$ & $33$ & $33$ \\
GPT-5            & $0$  & $0$  & $25$ & $75$ \\
GPT-5 Mini       & $0$  & $11$ & $67$ & $22$ \\
Gemini 2.5 Pro   & $0$  & $0$  & $45$ & $55$ \\
Gemini 2.5 Flash & $42$ & $0$  & $25$ & $33$ \\
\bottomrule
\end{tabular}
\caption{Per-model failure-mode distribution (\%) in the $n{=}64$ hand-labelled interactive-navigation sample. Per-model sample sizes: Opus~$12$, Haiku~$12$, GPT-5~$8$, GPT-5 Mini~$9$, Gemini~Pro~$11$, Gemini~Flash~$12$. Rows may not sum to $100\%$ due to rounding.}
\label{tab:per_model_failures}
\end{table}

\begin{table}[h]
\centering
\small
\setlength{\tabcolsep}{3.5pt}
\begin{tabular}{lrrrrrr}
\toprule
Scenario type & $n$ & FP & C1 & C2 & C3 & C4 \\
\midrule
page existence      & $120$ & $6\%$  & $42\%$ & $52\%$ & $0\%$  & $0\%$  \\
parent--child nav.\ & $189$ & $1\%$  & $21\%$ & $30\%$ & $15\%$ & $33\%$ \\
tab navigation      & $392$ & $10\%$ & $34\%$ & $48\%$ & $5\%$  & $3\%$  \\
\bottomrule
\end{tabular}
\caption{Human-label distribution (\%) among auto-flagged cases by scenario type, projected from the hand-labelled sample to the population. FP denotes false positives under the automatic evaluator. The decomposition matches the structural meaning of the three types: \emph{page-existence} failures concentrate on unreachable targets (C1, blank page) and route mapping errors (C2); \emph{parent--child} failures concentrate on inert affordances (C4) and missing targets (C3), which form the operability axis that state-isolated testing was designed to expose; \emph{tab-navigation} failures pattern with the page-existence axis. Percentages are rounded to the nearest integer.}
\label{tab:per_type_failures}
\end{table}

\section{Qualitative Failure Case Gallery}
\label{app:failure_gallery}

We illustrate the failure modes of \S\ref{sec:analysis} with hand-picked cases from the labelled pool. Each strip reads, left to right: reference source page, agent's rendered source page \emph{before} the trigger, agent's rendered page \emph{after} the trigger, and reference target page.

\begin{figure*}[h]
\centering
\includegraphics[width=\linewidth]{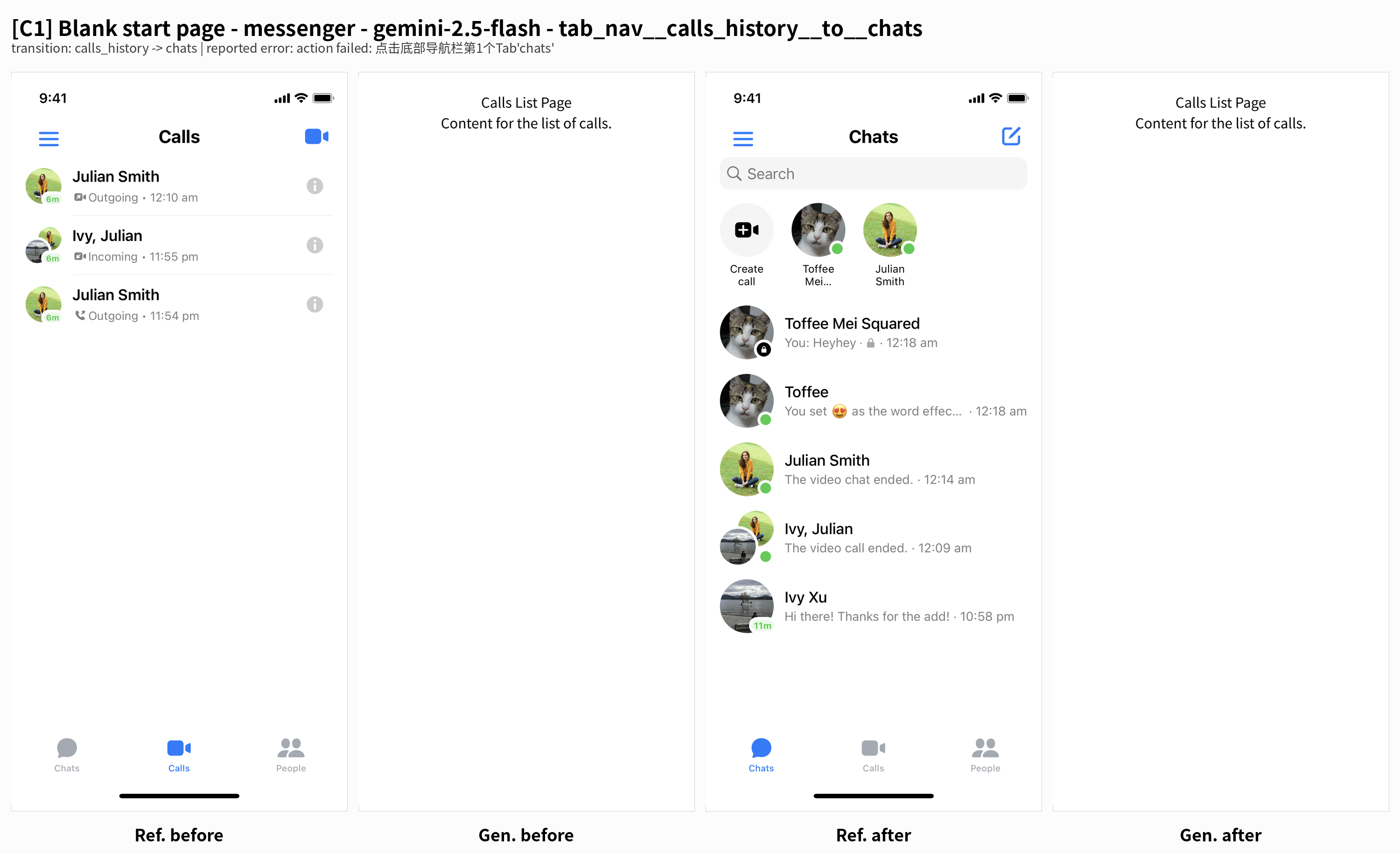}\\[2pt]
\includegraphics[width=\linewidth]{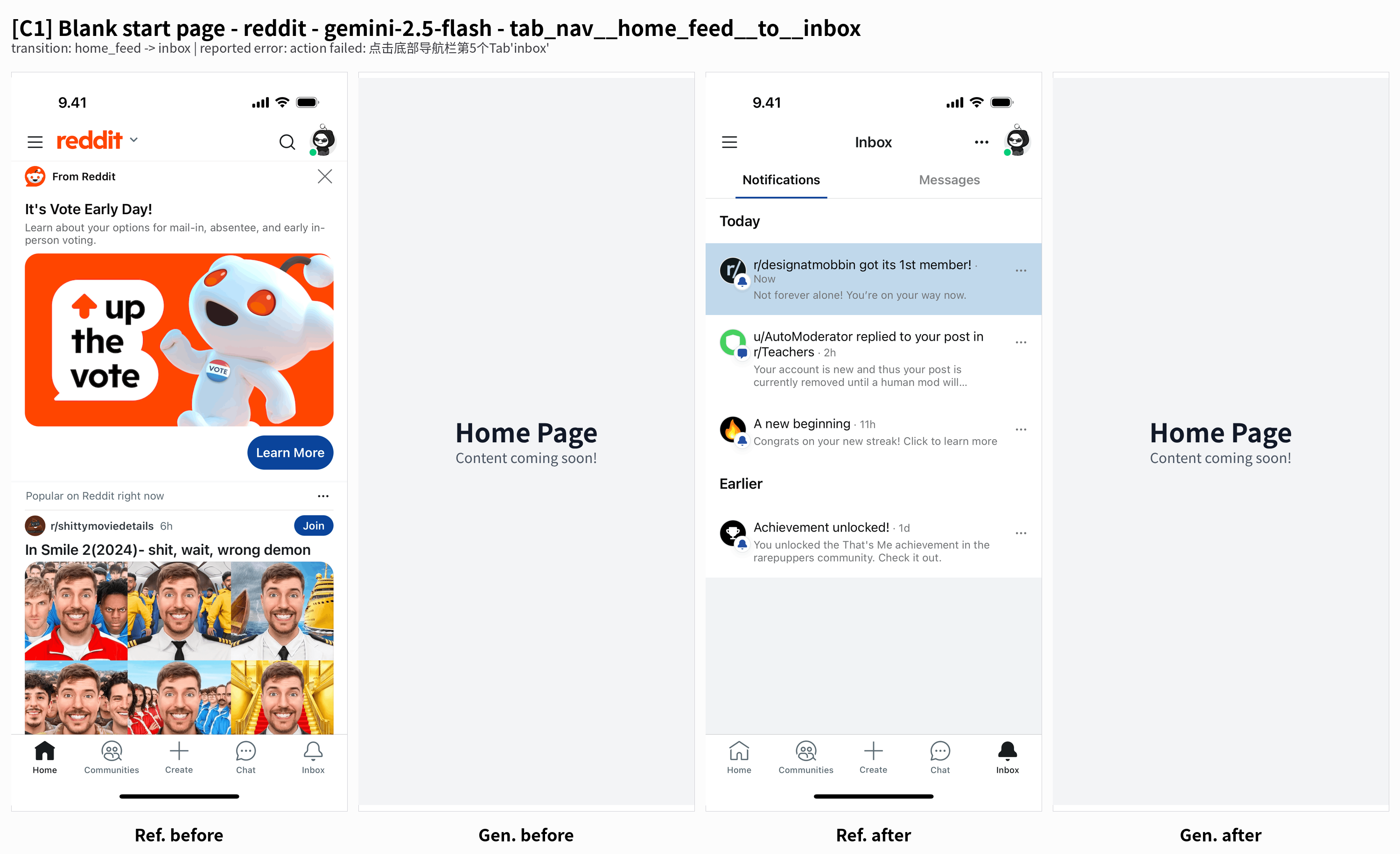}
\caption{\textbf{C1}: blank starting page (1/2). (a)~Messenger / Gemini~2.5~Flash, \texttt{calls\_history}~$\to$~\texttt{chats}; (b)~Reddit / Gemini~2.5~Flash, \texttt{home\_feed}~$\to$~\texttt{inbox}. The source page renders empty or as a stub placeholder with no interaction target, so no navigation action can be dispatched (\S\ref{sec:analysis}).}
\label{fig:case_C1_1}
\end{figure*}

\begin{figure*}[h]
\centering
\includegraphics[width=\linewidth]{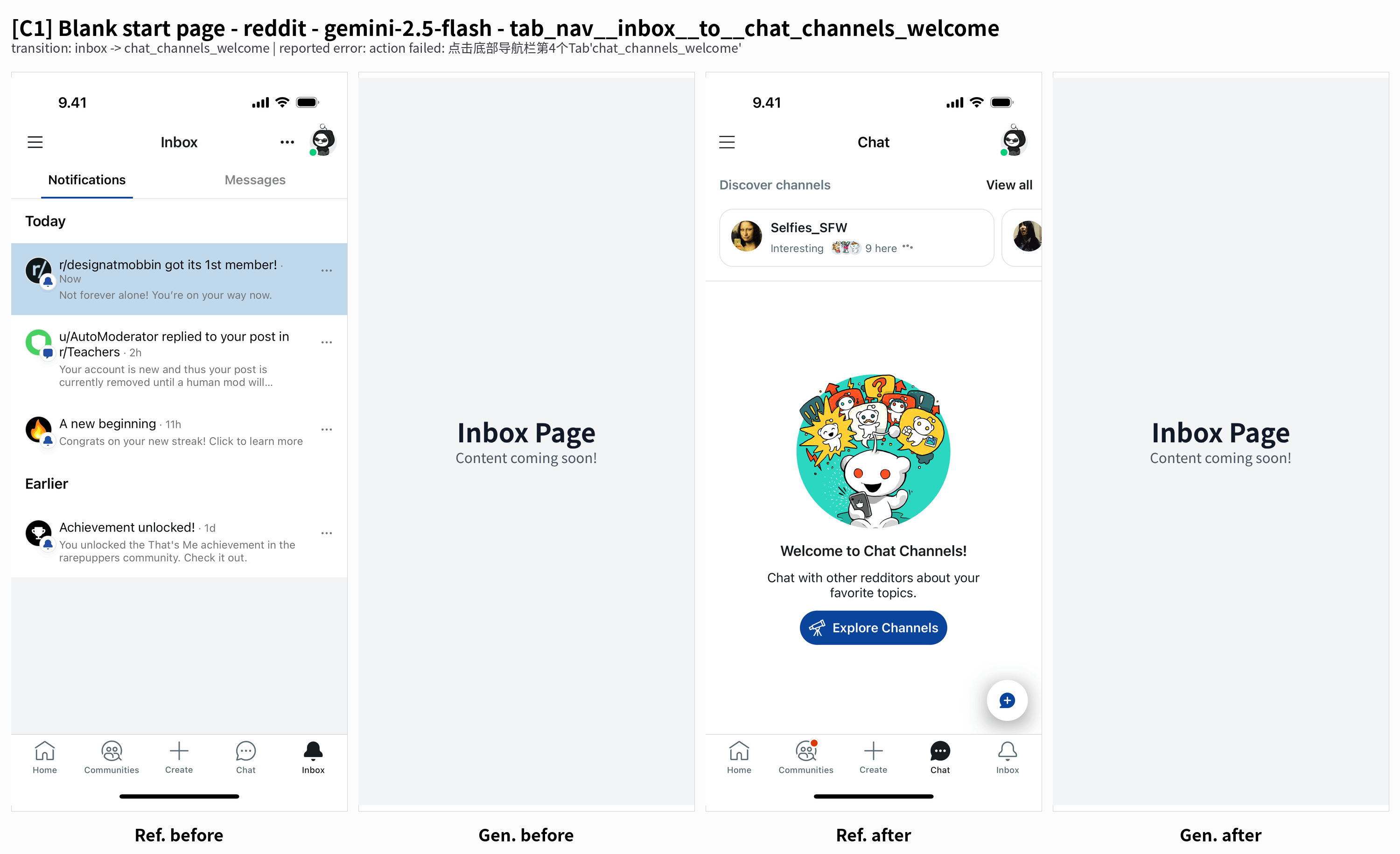}
\caption{\textbf{C1}: blank starting page (2/2). (c)~Reddit / Gemini~2.5~Flash, \texttt{inbox}~$\to$~\texttt{chat\_channels\_welcome}.}
\label{fig:case_C1_2}
\end{figure*}

\begin{figure*}[h]
\centering
\includegraphics[width=\linewidth]{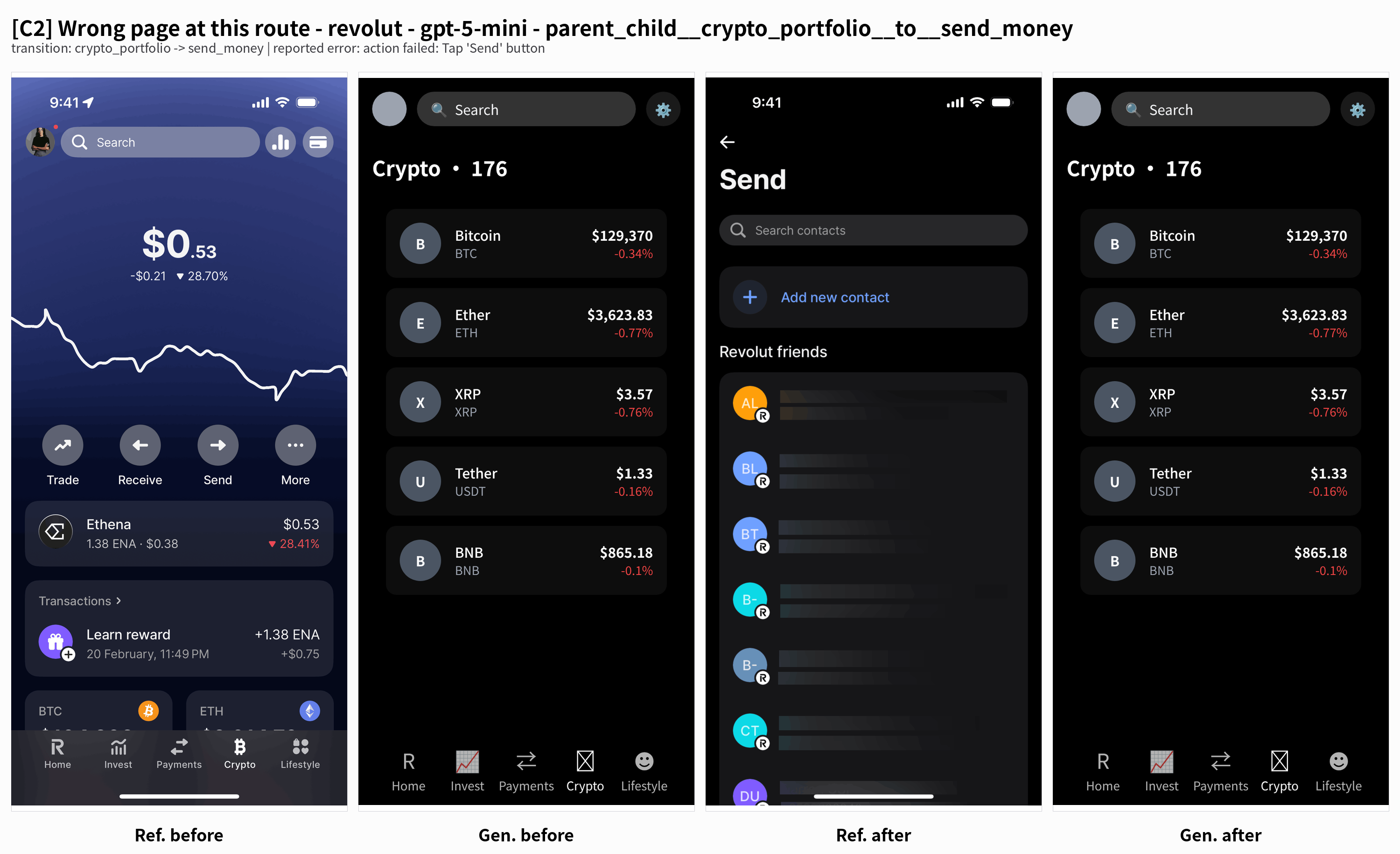}\\[2pt]
\includegraphics[width=\linewidth]{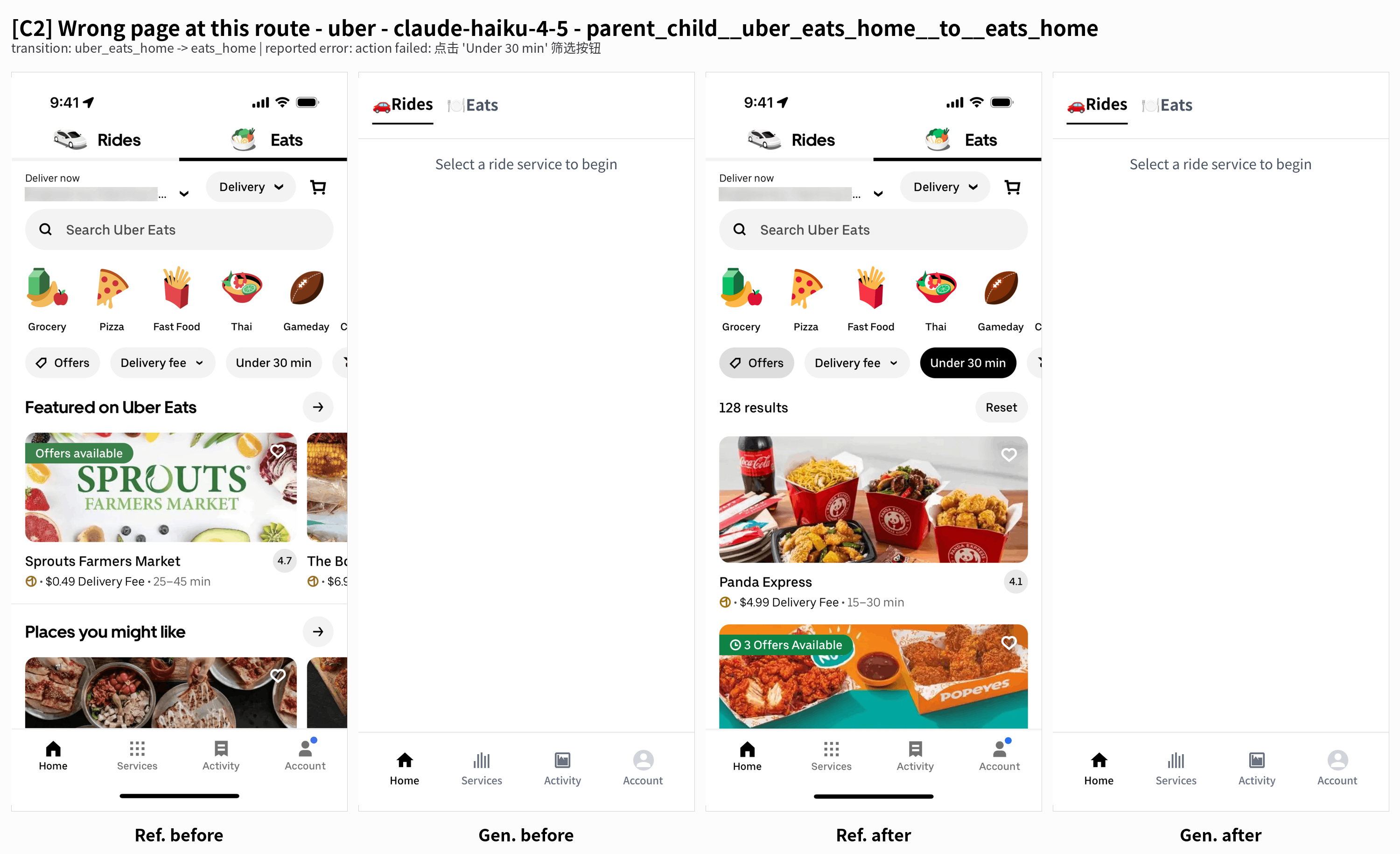}
\caption{\textbf{C2}: route mapping error. (a)~Revolut / GPT-5 Mini, \texttt{crypto\_portfolio}~$\to$~\texttt{send\_money}; (b)~Uber / Claude Haiku~4.5, \texttt{uber\_eats\_home}~$\to$~\texttt{eats\_home}. The agent declared the route, but the URL resolves to a different page than the one referenced by the test; the model has correctly produced the page, but the global routing table maps the navigation action to the wrong page (\S\ref{sec:analysis}).}
\label{fig:case_C2}
\end{figure*}

\begin{figure*}[h]
\centering
\includegraphics[width=\linewidth]{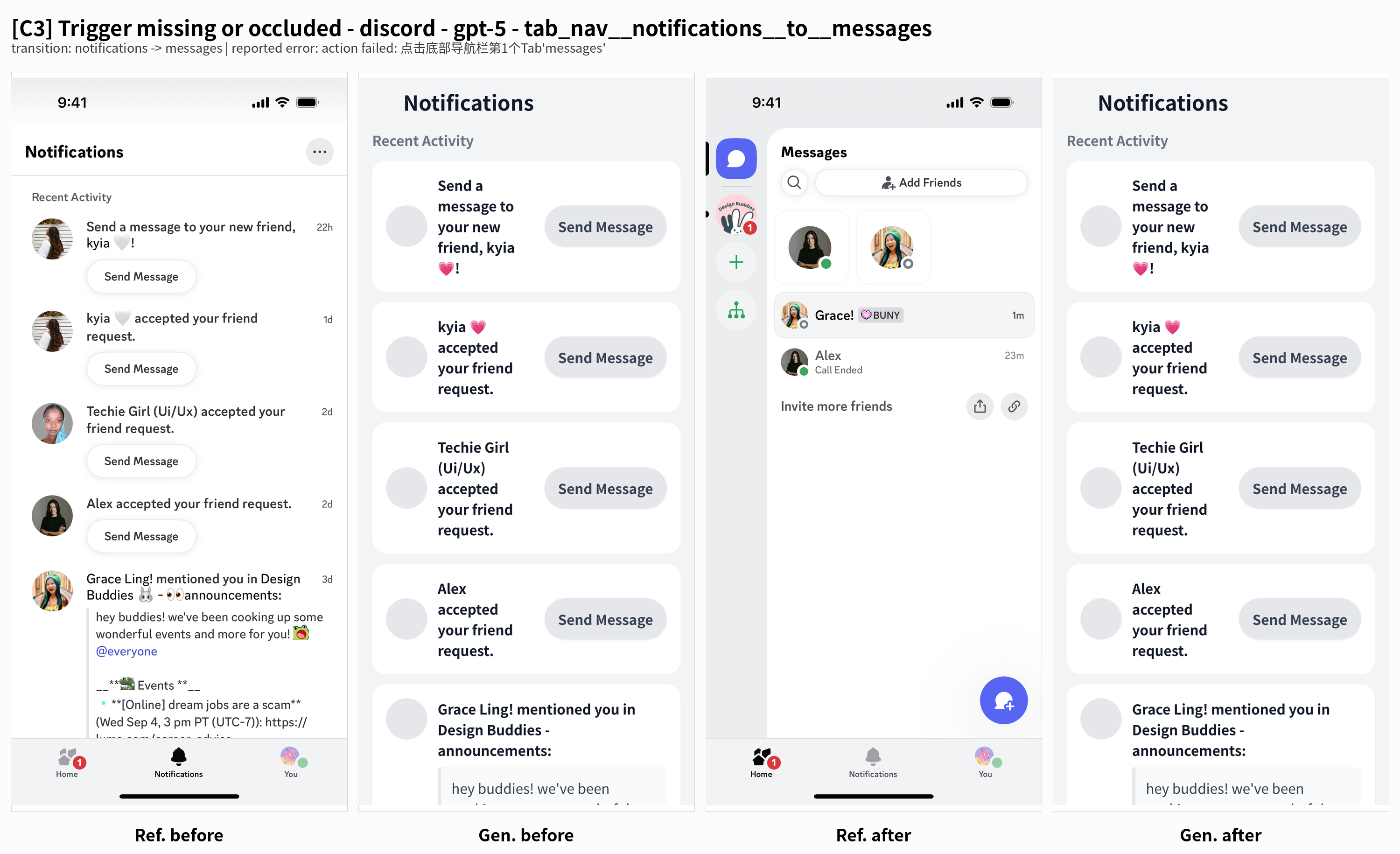}\\[2pt]
\includegraphics[width=\linewidth]{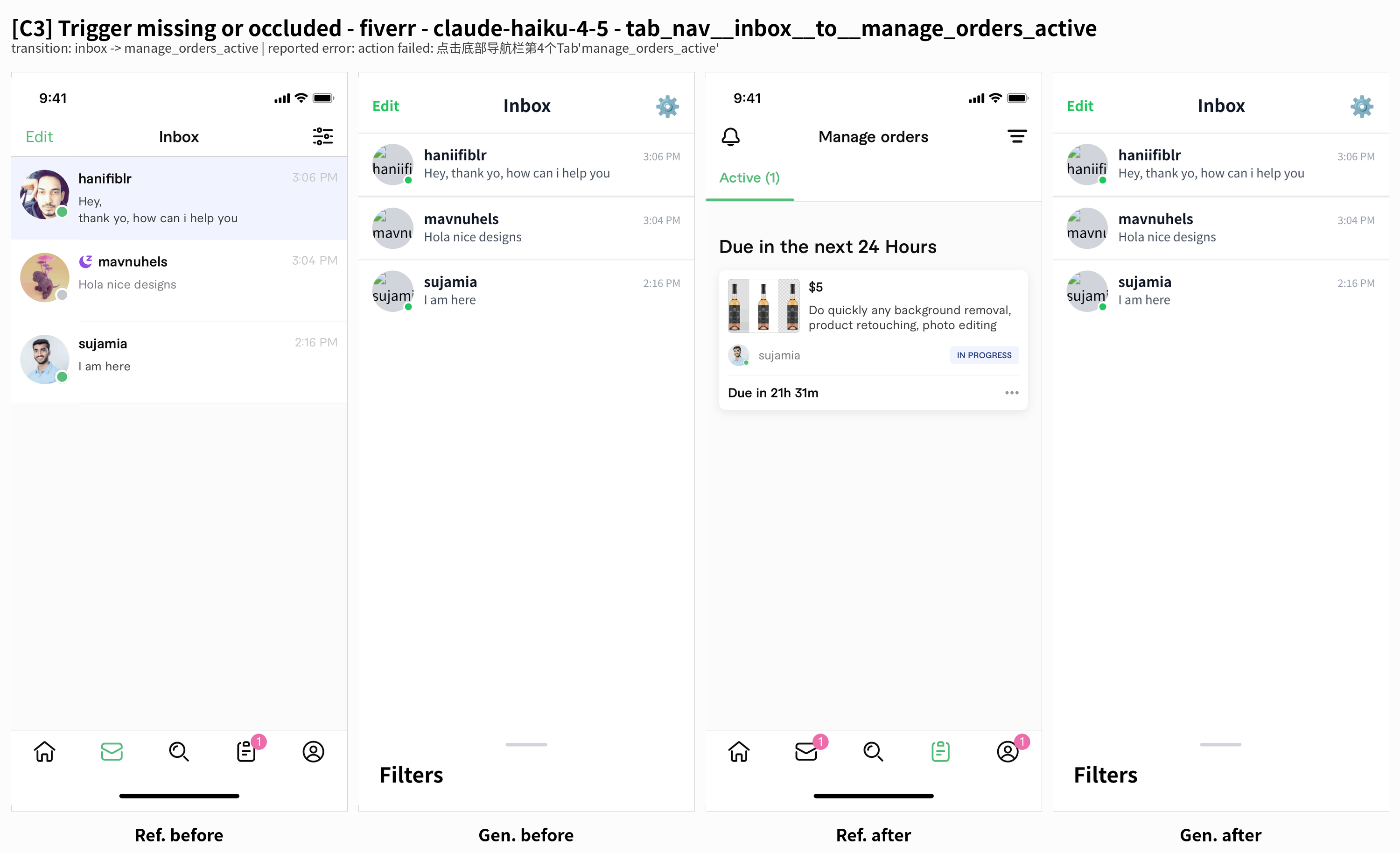}
\caption{\textbf{C3}: target unreachable or occluded (1/3). (a)~Discord / GPT-5, \texttt{notifications}~$\to$~\texttt{messages}; (b)~Fiverr / Claude Haiku~4.5, \texttt{inbox}~$\to$~\texttt{manage\_orders\_active}. The action target is not present in the rendered viewport, either because the layout omits it entirely or because another element covers it. This is structurally distinct from C4: here the page is malformed, whereas in C4 it is well-formed but inert (\S\ref{sec:analysis}).}
\label{fig:case_C3_1}
\end{figure*}

\begin{figure*}[h]
\centering
\includegraphics[width=\linewidth]{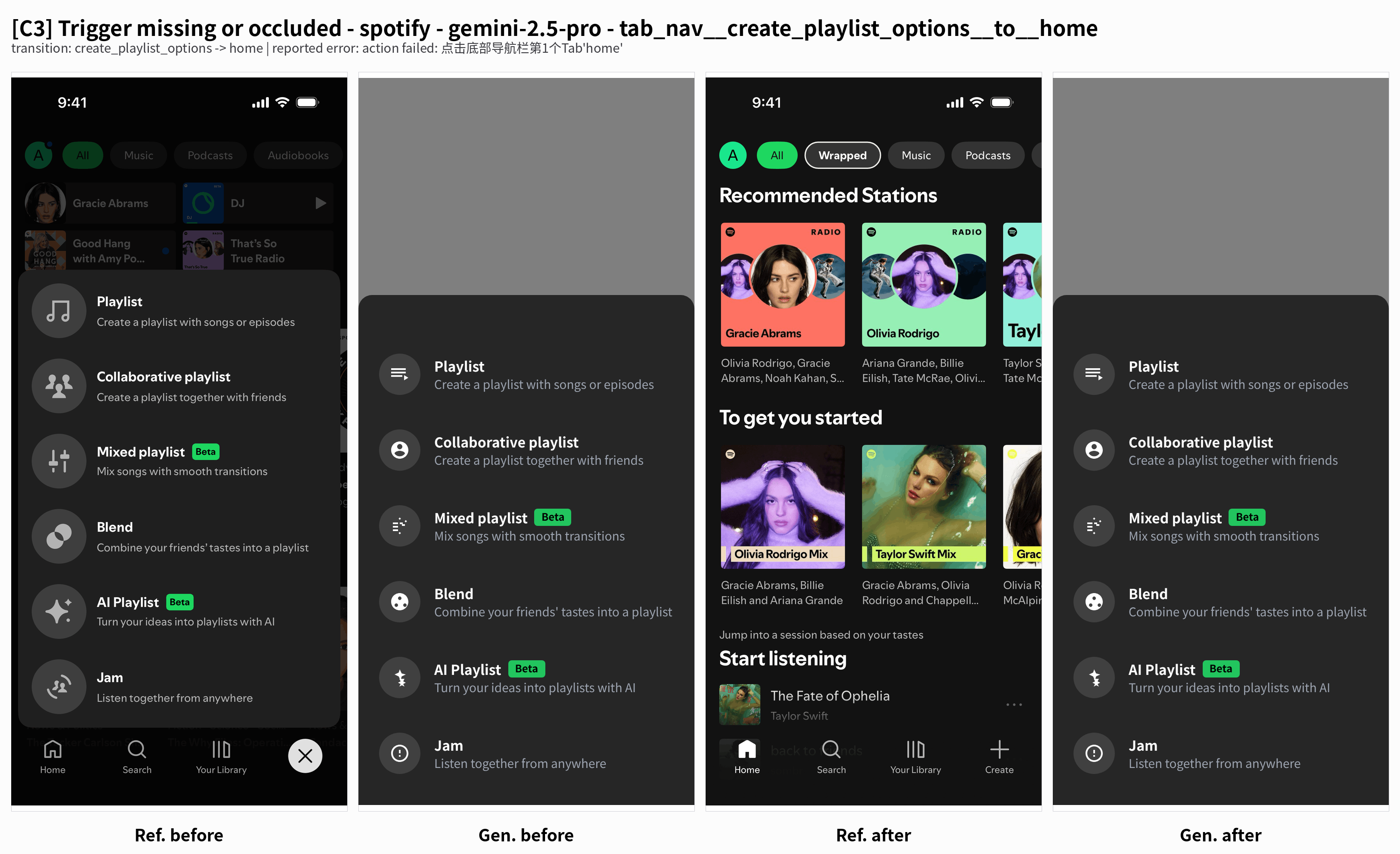}\\[2pt]
\includegraphics[width=\linewidth]{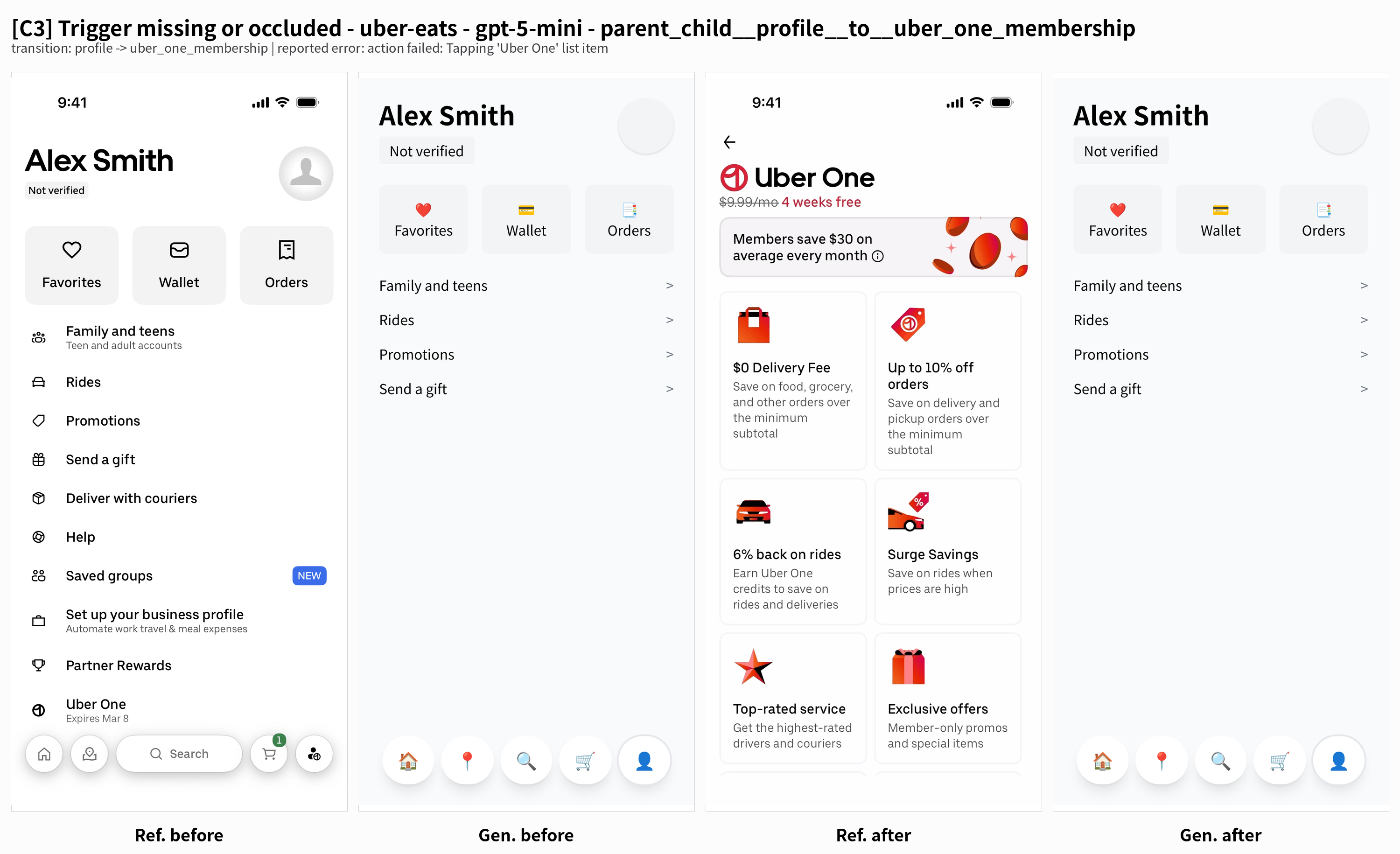}
\caption{\textbf{C3}: target unreachable or occluded (2/3). (c)~Spotify / Gemini~2.5~Pro, \texttt{create\_playlist\_options}~$\to$~\texttt{home}; (d)~Uber Eats / GPT-5 Mini, \texttt{profile}~$\to$~\texttt{uber\_one\_membership}.}
\label{fig:case_C3_2}
\end{figure*}

\begin{figure*}[h]
\centering
\includegraphics[width=\linewidth]{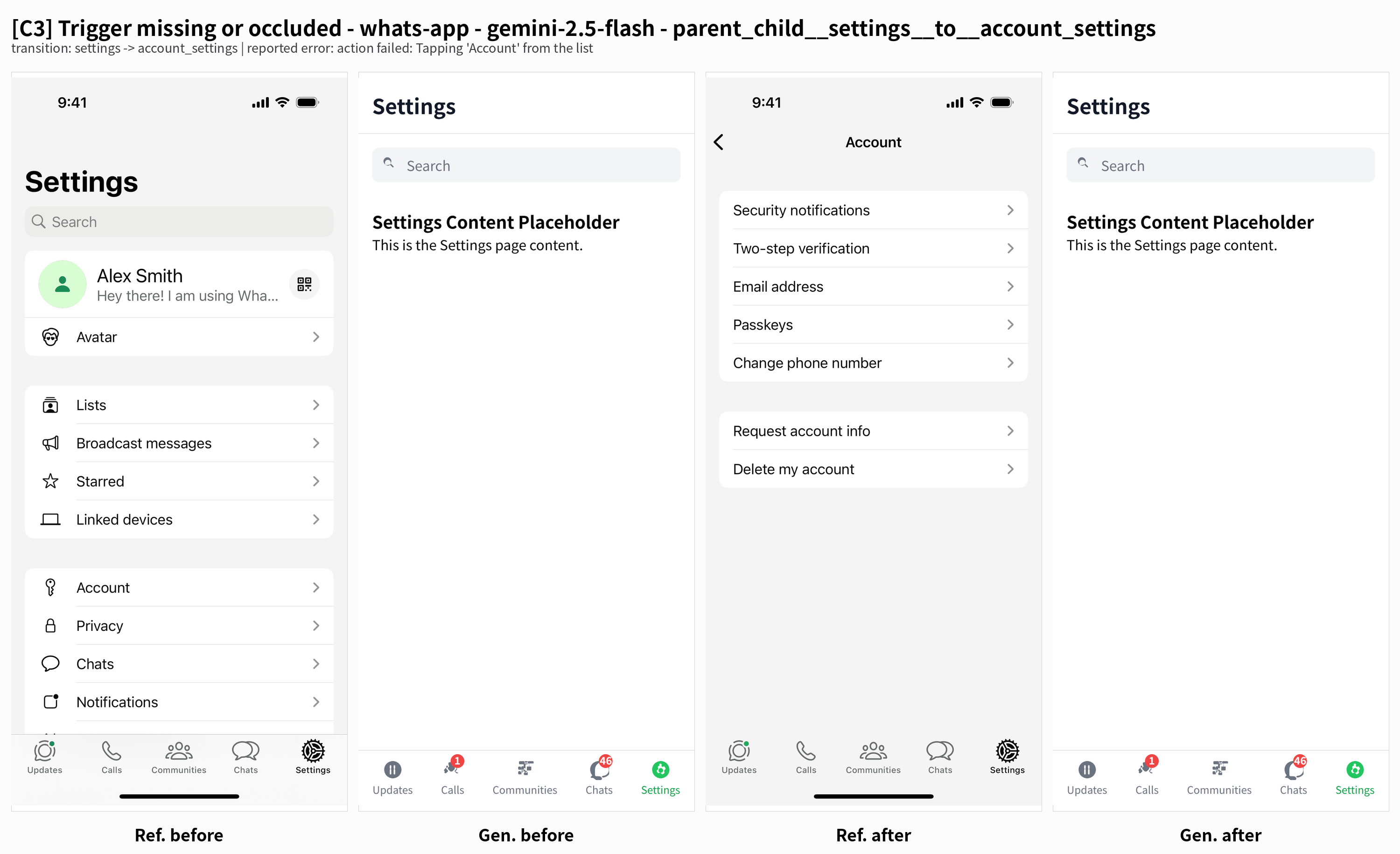}
\caption{\textbf{C3}: target unreachable or occluded (3/3). (e)~WhatsApp / Gemini~2.5~Flash, \texttt{settings}~$\to$~\texttt{account\_settings}.}
\label{fig:case_C3_3}
\end{figure*}

\begin{figure*}[h]
\centering
\includegraphics[width=\linewidth]{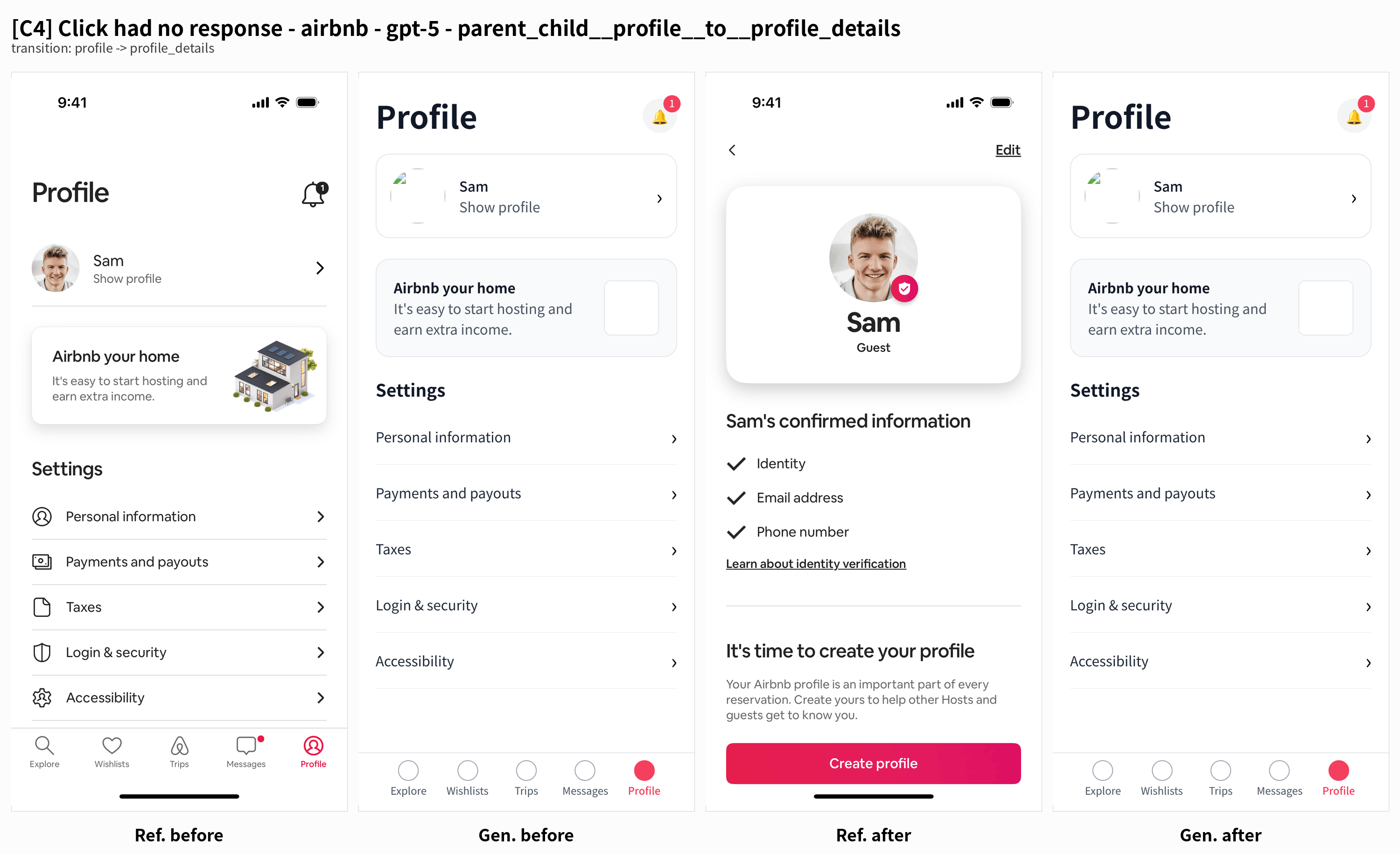}\\[2pt]
\includegraphics[width=\linewidth]{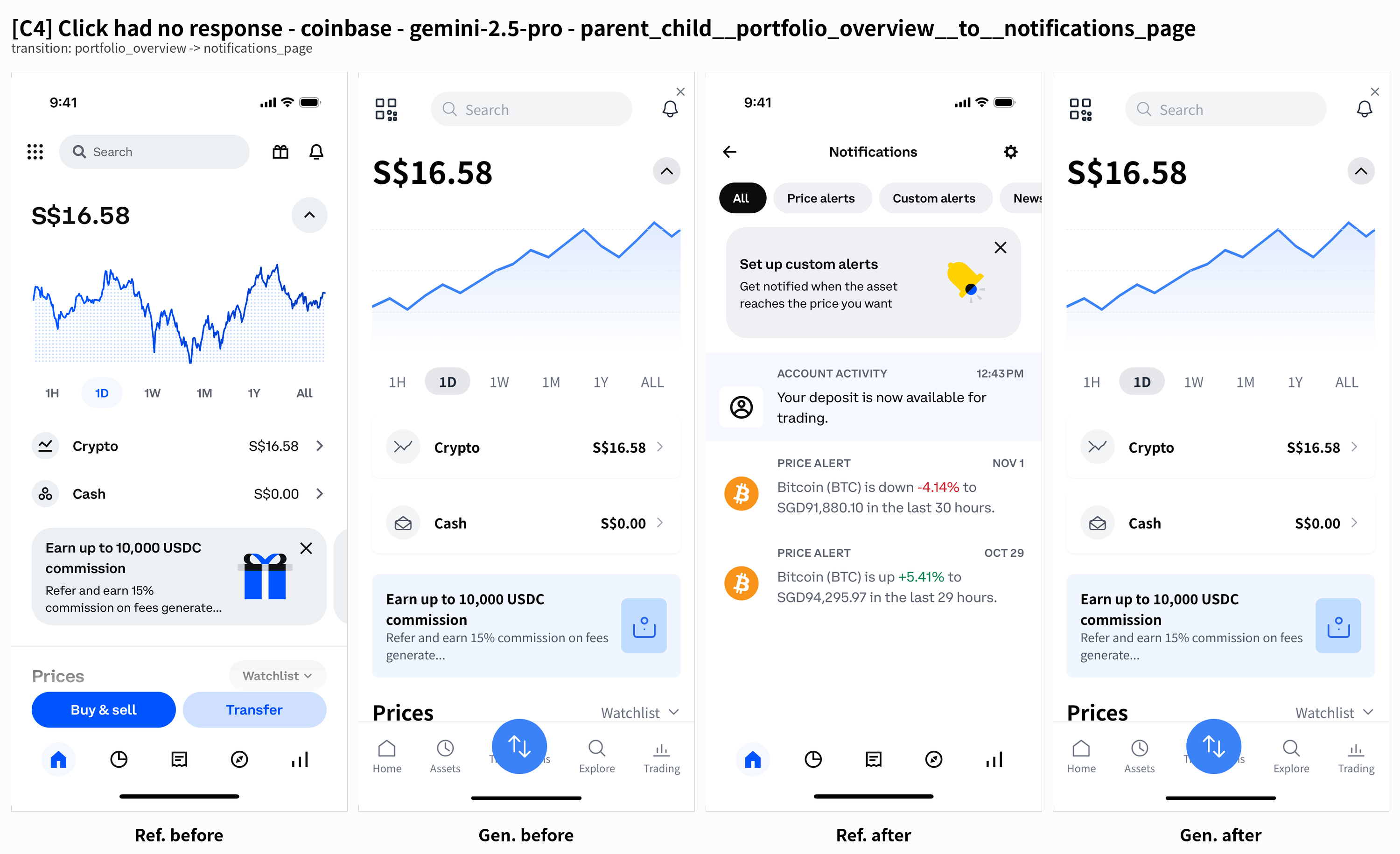}
\caption{\textbf{C4}: target clickable but unresponsive (1/3). (a)~Airbnb / GPT-5, \texttt{profile}~$\to$~\texttt{profile\_details}; (b)~Coinbase / Gemini~2.5~Pro, \texttt{portfolio\_overview}~$\to$~\texttt{notifications\_page}. The page renders, the affordance is in the right place, but the click registers no effect, consistent with a missing or mis-wired \texttt{onClick} handler (\S\ref{sec:analysis}).}
\label{fig:case_C4_1}
\end{figure*}

\begin{figure*}[h]
\centering
\includegraphics[width=\linewidth]{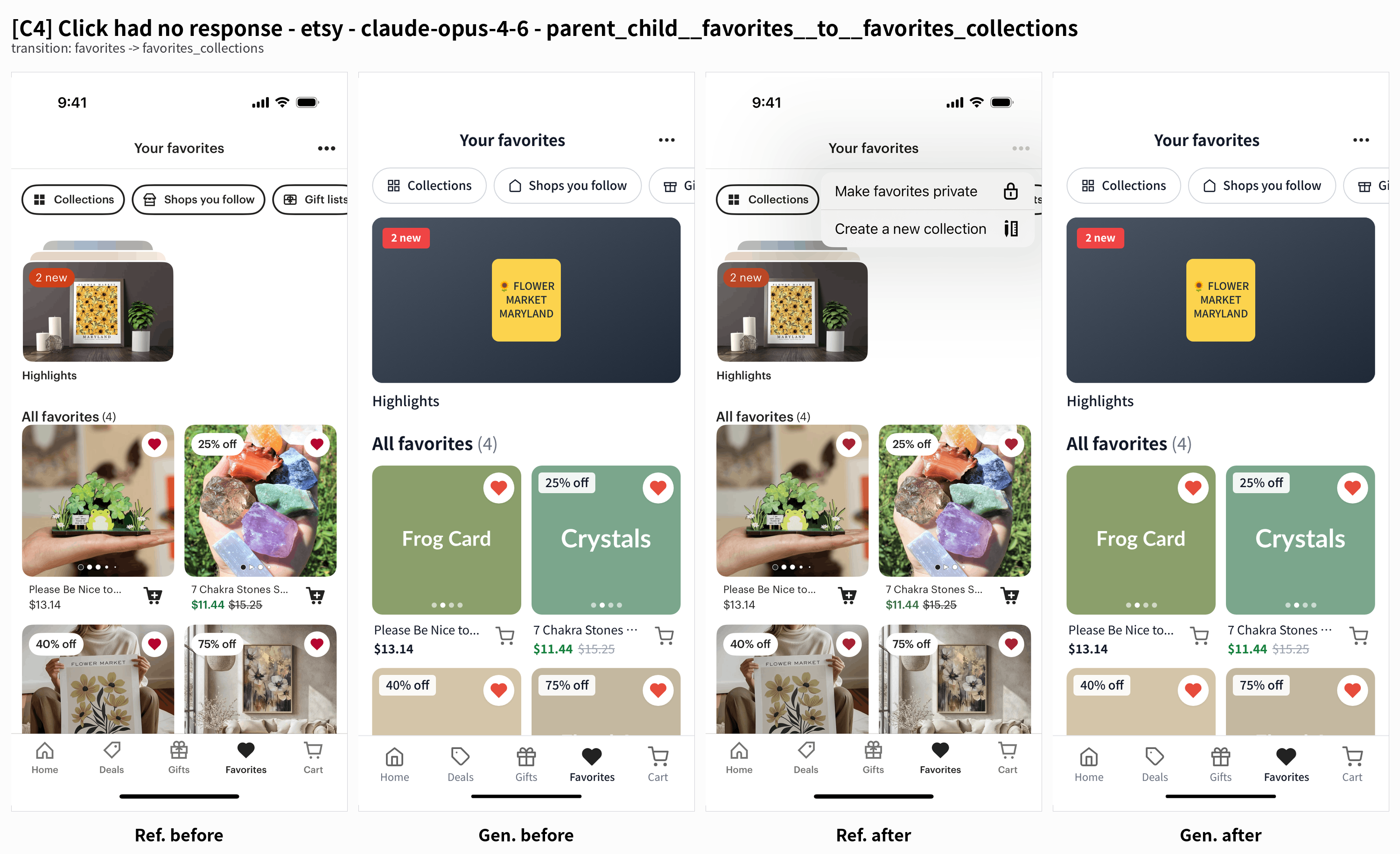}\\[2pt]
\includegraphics[width=\linewidth]{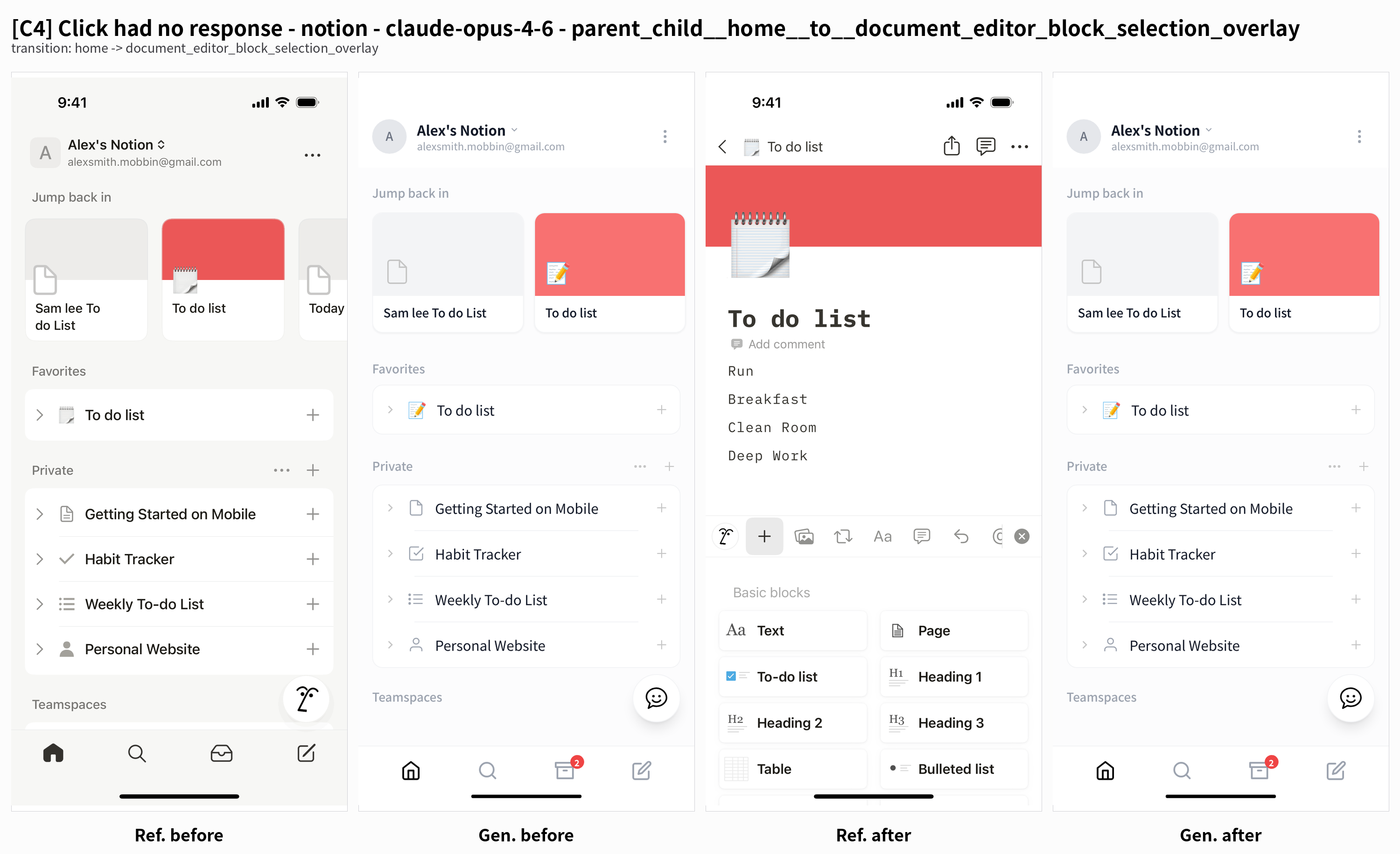}
\caption{\textbf{C4}: target clickable but unresponsive (2/3). (c)~Etsy / Claude Opus~4.6, \texttt{favorites}~$\to$~\texttt{favorites\_collections}; (d)~Notion / Claude Opus~4.6, \texttt{home}~$\to$~\texttt{document\_editor\_block\_selection\_overlay}.}
\label{fig:case_C4_2}
\end{figure*}

\begin{figure*}[h]
\centering
\includegraphics[width=\linewidth]{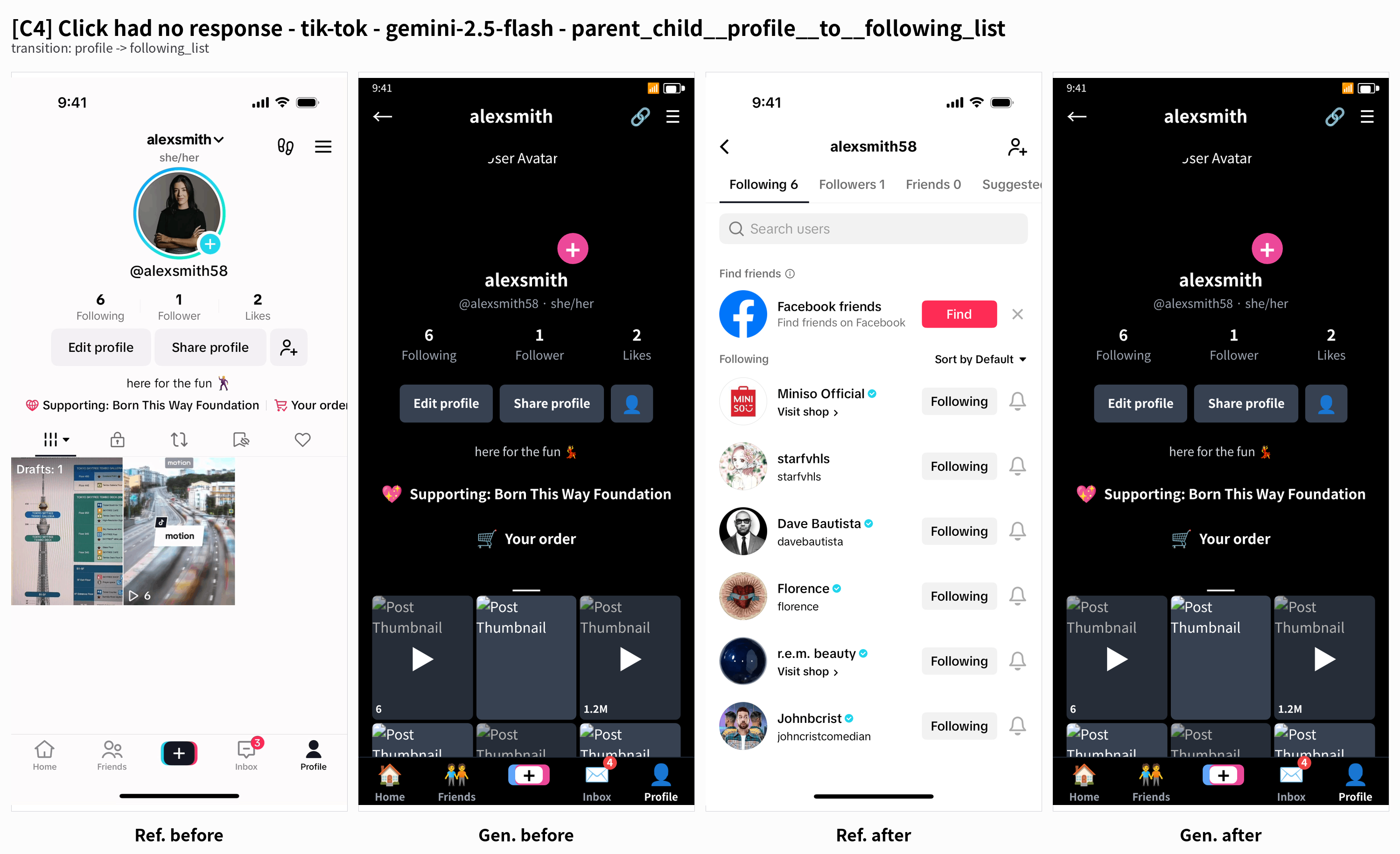}
\caption{\textbf{C4}: target clickable but unresponsive (3/3). (e)~TikTok / Gemini~2.5~Flash, \texttt{profile}~$\to$~\texttt{following\_list}.}
\label{fig:case_C4_3}
\end{figure*}

\section{Code Maintainability Detail}
\label{app:engineering}

Table~\ref{tab:eng_full} extends Table~\ref{tab:maintainability} with the remaining per-model code maintainability indicators: total file count, shared-component count, and standard deviations across apps. As noted in \S\ref{sec:engineering}, design-token color consistency has saturated across all frontier models ($0.74$--$0.85$, $15\%$ spread), while structural indicators (LoC, LoC/file, reuse, dead-component rate) differentiate sharply.

\begin{table}[h]
\centering
\small
\begin{tabular}{lrrr}
\toprule
Model & Files & Shared comp. & LoC s.d. \\
\midrule
Claude Opus 4.6  & $17.6$ & $4.0$ & $325$ \\
Claude Haiku 4.5 & $18.9$ & $5.6$ & $382$ \\
GPT-5            & $20.1$ & $5.4$ & $189$ \\
GPT-5 Mini       & $16.7$ & $4.3$ & $146$ \\
Gemini 2.5 Pro   & $22.8$ & $3.5$ & $213$ \\
Gemini 2.5 Flash & $22.9$ & $9.2$ & $276$ \\
\bottomrule
\end{tabular}
\caption{Additional code maintainability indicators per model (means over $29$ apps). \texttt{Files}: number of source files. \texttt{Shared comp.}: count of components under \texttt{src/components/}. \texttt{LoC s.d.}: standard deviation of LoC across apps for this model.}
\label{tab:eng_full}
\end{table}

\end{document}